\documentclass[%
 reprint,
 amsmath,amssymb,
 aps,
]{revtex4-2}

\usepackage{physics,graphicx,xcolor}% Include figure files
\usepackage{colortbl}
\usepackage{multirow}
\usepackage{setspace}
\usepackage{dsfont}

\newcommand{\cF}{\mathcal{F}}
\newcommand{\cR}{\mathcal{R}}
\newcommand{\iL}{\mathcal{L}^{-1}}
\newcommand{\tR}{\widetilde{R}}
\newcommand{\tS}{\widetilde{S}}
\newcommand{\tcR}{\widetilde{\mathcal{R}}}
\newcommand{\1}{\mathds{1}}
\newcommand{\bS}{{\bf S}}
\graphicspath{ {./Figure/} }
\begin{document}

%\preprint{APS/123-QED}

\title{Random density matrices: Analytical results for mean fidelity and variance of squared Bures distance}% Force line breaks with \\
%\thanks{A footnote to the article title}%

\author{Aritra Laha}
\author{Santosh Kumar}
 \email{skumar.physics@gmail.com}
\affiliation{%
 Department of Physics, Shiv Nadar Institution of Eminence,
Gautam Buddha Nadar, Uttar Pradesh 201314, India
}%

%\date{\today}% It is always \today, today,
             %  but any date may be explicitly specified

\begin{abstract}
One of the key issues in quantum information theory related problems concerns with that of distinguishability of quantum states. In this context, Bures distance serves as one of the foremost choices among various distance measures. It also relates to fidelity, which is another quantity of immense importance in quantum information theory. In this work, we derive exact results for the average fidelity and variance of the squared Bures distance between a fixed density matrix and a random density matrix, and also between two independent random density matrices. These results supplement the recently obtained results for the mean root fidelity and mean of squared Bures distance [Phys. Rev. A {\bf 104}, 022438 (2021)]. The availability of both mean and variance also enables us to provide a gamma-distribution-based approximation for the probability density of the squared Bures distance. The analytical results are corroborated using Monte Carlo simulations. Furthermore, we compare our analytical results with the mean and variance of the squared Bures distance between reduced density matrices generated using coupled kicked tops, and a correlated spin chain system in a random magnetic field. In both cases, we find good agreement.   
%\begin{description}
%\item[Usage]
%Secondary publications and information retrieval purposes.
%\item[PACS numbers]
%May be entered using the \verb+\pacs{#1}+ command.
%\item[Structure]
%You may use the \texttt{description} environment to structure your abstract;
%use the optional argument of the \verb+\item+ command to give the category of each item. 
%\end{description}
\end{abstract}

\pacs{Valid PACS appear here}% PACS, the Physics and Astronomy
                             % Classification Scheme.
%\keywords{Suggested keywords}%Use showkeys class option if keyword
                              %display desired
\maketitle

%\tableofcontents

%%%%%%
\section{Introduction}
%%%%%%

Our modern world is driven by technology and information theory plays a central role in it. It serves as the basis of all communications, networking, and data storage systems. While currently most of these tasks rely on classical means, gradually we are aiming towards employing quantum mechanical means to accomplish information-processing tasks~\cite{NC2000,E2009,BZ2017,W2017}. Quantum information theory deals with the study of such information processing tasks and their achievable limits within quantum mechanics. In this context, various distance measures between quantum states play an important role and are useful in diverse problems related to quantum communication protocols, quantification of quantum correlations, and quantum-state tomography,  etc~\cite{NC2000,E2009,BZ2017,W2017,STM2013,KKF2020,ZE2011,TBCL2019,Gao2016,VPRK1997,ACSZC2019,LZ2004,SO2013,WPSW2020,PSW2020,LTOCC2019,KLPCSC2019,CPCC2019,RSI2016,MMPZ2008,PPZ2016}. The distance measure focussed upon in this work is the Bures distance, which is Riemannian and also monotone, where the latter feature implies that it does not grow under any completely positive trace preserving quantum operation. Due to this reason, it is one of the foremost choices in quantifying distinguishability of quantum states~\cite{RSI2016,MMPZ2008,PPZ2016} and finds applications in the problems listed above. 

Consider two quantum states represented by the density matrices $\rho_1$ and $\rho_2$, respectively. The Bures distance is then defined as~\cite{BZ2017,W2017,B1969,U1976},
\begin{align}
\label{bd}
d_{B}(\rho_1,\rho_2)=\sqrt{2-2\sqrt{\mathcal{F}(\rho_1,\rho_2)}},
\end{align}
where $\mathcal{F}(\rho_1,\rho_2)$ is the fidelity between the two states and is given by~\cite{B1969,U1976,U1986,J1994},
\begin{align}
\label{fid}
\mathcal{F}(\rho_1,\rho_2)=\left(\tr \sqrt{\sqrt{\rho_1}\rho_2 \sqrt{\rho_1}}\right)^2,
\end{align}
The notation $`\tr$' in the above equation represents the trace. The Bures distance takes the minimum value of 0 when the two states $\rho_1$ and $\rho_2$ are identical, whereas the maximum value of $\sqrt{2}$ is achieved when the two states are supported on orthogonal subspaces.

To examine various properties of quantum states under investigation in quantum information theory related problems, some reference states are frequently needed. Random quantum states constitute a natural choice in this context since they provide the most typical or generic characteristic. They are also useful in describing noise-affected quantum states~\cite{W1990,H1998,ZS2001,SZ2004,ZPNC2011,CN2016}. Additionally, they play a crucial role in the area of quantum chaos where one examines quantum states whose classical analog is chaotic~\cite{W1990,HKS1987,H2010}.  For random pure states in a finite dimensional Hilbert space, Haar measure on the unitary group serves as a natural probability measure. However, as far as random mixed states are concerned there is no such unique measure to describe their statistics~\cite{W1990,H1998,ZS2001,SZ2004}. A widely used probability measure is the Hilbert-Schmidt measure on the set of the finite-dimensional mixed-sates~\cite{H1998,ZS2001,ZS2001,SZ2004,ZPNC2011,CN2016,Lubkin1978,LP1988,Page1993,G2007,MBL2008,NMV2011,
KP2011,VPO2016,KSA2017,K2019,FK2019,W2020}. Statistical investigation of various distance measures involving these Hilbert-Schmidt random states has become an active area of research due to its fundamental as well as applied aspects~\cite{PPZ2016,B1996,ZS2005,HLW2006,M2007,BSZW2016,MZB2017,K2020,KC2020,LAK2021}. 

Recently, exact analytical results for the mean-square Hilbert-Schmidt and Bures distances have been derived in Refs.~\cite{K2020,LAK2021} between a fixed density matrix and a random density matrix and also between two random density matrices. In the present work, we go beyond the mean and obtain exact analytical expressions for the variance of the squared Bures distance, which is given by,
\begin{align}
\label{b2}
D(\rho_1,\rho_2)\equiv d_{B}^{2}(\rho_1,\rho_2)=2-2\sqrt{\mathcal{F}(\rho_1,\rho_2)}.
\end{align}
Its mean and variance can be calculated as,
\begin{align}
\label{mea}
\expval{D}=2-2\expval{\sqrt{\cF}},
\end{align}
\begin{align}
\label{var}
\nonumber
\mathrm{Var}(D)&=\langle(2-2\sqrt{\mathcal{F}})^2\rangle-\langle 2-2\sqrt{\mathcal{F}}\rangle^2\\
~~~~~&=4(\langle\mathcal{F}\rangle-\langle\sqrt{\mathcal{F}}\rangle^2),
\end{align}
where $\cF$ is the fidelity as defined in Eq.~\eqref{fid}. The averaging $\langle .\rangle$ in the above expressions is with respect to the probability measure followed by the random density matrices involved in the calculation. Exact expressions for the mean root fidelity $\langle\sqrt{\mathcal{F}}\rangle$ and hence the mean-square Bures distance are already known from our earlier work~\cite{LAK2021}. Therefore, our main task in this work is to obtain analytical expressions for the mean fidelity, i.e., $\langle\mathcal{F}\rangle$. For this task, we utilize the well-established connection between the Hilbert-Schmidt and Wishart-Laguerre ensemble of random matrices. With the availability of both the mean and variance, we are also able to provide a Gamma distribution based approximation to the probability distribution function (PDF) of the squared Bures distance.

We organize the rest of the paper as follows.  In Sec.~\ref{sec2} we derive the exact expressions for the average fidelity and hence the variance of the squared Bures distance. In Sec.~\ref{sec3} we propose a gamma-distribution based approximation for the PDF of squared Bures distance using the cumulant-matching approach. In these two sections, we also validate our analytical results by comparing them with Monte Carlo simulations based on relevant matrix models.  In Sec.~\ref{sec4} we compare our analytical results for squared Bures distance with those derived from coupled-kicked-top system and a random spin chain model. Sec. ~\ref{sec5} is dedicated to the summary and outlook of our work. Details of some of the derivations are provided in the Appendixes.

%%%%%%
\section{Average fidelity and variance of the squared Bures distance between random density matrices} 
\label{sec2}
%%%%%%

For the sake of completeness, we begin by recalling some basic concepts related to the construction of Hilbert-Schmidt mixed state in the bipartite framework. Consider a bipartite system comprising two subsystems described by $n$-dimensional and $m$-dimensional Hilbert spaces $\mathcal{H}_{n}$ and $\mathcal{H}_{m}$, respectively. The composite bipartite system is then described by the $nm$-dimensional Hilbert space $\mathcal{H}_{n}\otimes \mathcal{H}_{m}$. We take $n\le m$ without loss of any generality. Now, consider an arbitrary state $|\psi_{0}\rangle$, taken from $\mathcal{H}_n\otimes\mathcal{H}_m$, then a random bipartite pure state $|\psi\rangle$ can be constructed as, $|\psi\rangle=U_{nm}|\psi_{0}\rangle$, where $U_{nm}$ is a global unitary operator taken from the Haar measure. We now perform the operation of partial tracing on this pure state, with respect to the $m$-dimensional subsystem, to obtain a reduced random state $\rho$, i.e.,
\begin{equation}
\label{rho}
\rho=\frac{\tr_{m}(|\psi\rangle\langle\psi|)}{\langle\psi|\psi\rangle}.
\end{equation}
This reduced state, in density matrix representation, is $n$-dimensional and is described by the Hilbert-Schmidt probability measure, for which the probability density function (PDF) is given by~\cite{ZS2001},
\begin{equation}
\label{dist}
\mathcal{P}(\rho)=C(\det \rho)^{m-n}~\delta(\tr \rho-1)~\Theta(\rho).
\end{equation}
Here, `$\det$' represents determinant and, as mentioned above, `$\tr$' stand for trace.  The Dirac delta-function $\delta(.)$ in the above equation fixes the trace of the density matrix to unity, and the Heaviside theta function $\Theta(.)$ imposes the positive-definiteness condition. Further, the normalization factor $C$ in Eq.~\eqref{dist} is given by,
\begin{equation}
\label{norm}
C=\Gamma(nm)\left(\pi^{n(n-1)/2}~\prod_{j=1}^{n}\Gamma(m-j+1)\right)^{-1}.
\end{equation}
Within random matrix theory (RMT), this reduced density matrix $\rho$ can be obtained by normalizing a Wishart-Laguerre matrix $W$ by its trace~\cite{ZS2001,SZ2004}, viz.,
\begin{equation}
\label{map}
\rho=\frac{W}{\tr(W)},
\end{equation}
The random matrix $W$ follows (unconstrained) Wishart-Laguerre PDF,
\begin{equation}
\label{PW}
P(W)=(\Gamma(nm))^{-1}C\, (\det W)^{m-n}e^{-\tr W}\Theta(W).
\end{equation}
Consequently, $\rho$ is also referred to as a fixed (unit) trace Wishart-Laguerre matrix.

In what follows, we calculate the mean fidelity and eventually the variance of the squared Bures distance between two density matrices when one is fixed and one is random, and also when both are random and independent.

%%%%%%
\subsection{A fixed density matrix $\sigma$ and a random density matrix $\rho$}
%%%%%%

As seen above, to evaluate the variance of the squared Bures distance, we need to evaluate $\langle\cF\rangle$ and $\langle\sqrt{\cF}\rangle$. Following Ref.~\cite{LAK2021}, we know that these averages can be recast in terms of the averages involving the (unordered) eigenvalues $\{\lambda_i\}$ of the matrix $\sqrt{\sigma}\rho \sqrt{\sigma}$. Moreover, these eigenvalues are related to the (unordered) eigenvalues $\{x_i\}$ of the random matrix $\sqrt{\sigma}W \sqrt{\sigma}$, where $W$ is the Wishart-Laguerre random matrix as in the Eq.~\eqref{PW}. The product form $\sqrt{\sigma}W \sqrt{\sigma}$ gives rise to the semi-correlated Wishart ensemble with $\sigma$ as the covariance matrix. For our calculation, we need the first and second order marginal densities of eigenvalues, or equivalently, the one-point and two-point correlation functions. Representing the $r$th point correlation function for $\sqrt{\sigma}\rho \sqrt{\sigma}$ and $\sqrt{\sigma}W \sqrt{\sigma}$ respectively by $\cR_r(\lambda_1,...,\lambda_r)$ and $R_r(x_1,...,x_r)$, we have the following relationship for $r=1$ and $r=2$ (see, e.g., Ref.~\cite{LAK2021}),
\begin{equation}
\label{R1Laplace}
\cR_1(\lambda)=\Gamma(nm)\mathcal{L}^{-1}[s^{1-nm}R_1(s\lambda),\{s,t\}]_{t=1},
\end{equation} 
\begin{align}
\label{R2Laplace}
\cR_{2}(\lambda_1,\lambda_2)=\Gamma(nm)\mathcal{L}^{-1}\left[s^{2-nm}R_{2}(s\lambda_1,s\lambda_2),\{s,t\}\right]_{t=1},
\end{align}
with $s>0$.
The eigenvalue correlation functions $R_1(x)$ and $R_2(x_1,x_2)$ pertaining to the semicorrelated Wishart ensemble are already known~\cite{G1963,ATLV2004,SMM2006,RKG2010},
\begin{equation}
\label{R1x}
R_1(x)=S(x,x),
\end{equation}
\begin{equation}
\label{R2xy}
R_2(x_1,x_2)=R_1(x_1)R_1(x_2)-S(x_1,x_2)S(x_2,x_1),
\end{equation}
where the kernel $S(x_1,x_2)$ is given by
\begin{equation}
\label{ker}
S(x_1,x_2)= \frac{1}{\Delta(\{a\})}\sum_{i=1}^{n}\det[g_{j,k}^{(i)}(x_1,x_2)]_{j,k=1}^{n}.
\end{equation}
In the above expression, $\Delta(\{a\})=\prod_{r>l}(a_r-a_l)$ is the Vandermonde determinant and
\begin{align}
g^{(i)}_{j,k}(x_1,x_2)=\begin{cases}\frac{a_j^m x_1^{n-i}x_2^{m-n}~e^{-a_j x_2}\Theta(x_1)\Theta(x_2)}{\Gamma(m-i+1)}, & k=i,\\
 a_{j}^{k-1}, &  k\ne i.
\end{cases}
\end{align}
Here, $\{a_{i}^{-1}\}$ are the eigenvalues of the fixed density matrix $\sigma$.

We now proceed to calculate the mean fidelity $\expval{\cF}$. We have,
\begin{align}
\label{mF}
\nonumber
\langle\mathcal{F}\rangle&=\left\langle\left(\tr \sqrt{\sqrt{\sigma}\rho \sqrt{\sigma}}\right)^{2}\right\rangle=\left\langle\sum_{j,k=1}^{n}\lambda_{j}^{1/2}\lambda_{k}^{1/2}\right\rangle\\
&=\left\langle\sum_{j=1}^{n}\lambda_{j}\right\rangle+\left\langle\sum_{{j,k=1}\atop{(j\ne k)}}^{n}\lambda_{j}^{1/2}\lambda_{k}^{1/2}\right\rangle.
\end{align}
For the first term, we obtain
\begin{align}
\label{R1avg}
\nonumber
&\left\langle\sum_{j=1}^{n}\lambda_{j}\right\rangle=\int_{-\infty}^{\infty} \lambda\cR_1(\lambda)d\lambda\\
\nonumber
&=\Gamma(nm)\iL\Big[s^{1-nm}\int_{-\infty}^{\infty} \lambda R_1(s\lambda)d\lambda,\{s,t\}\Big]_{t=1}\\
\nonumber
&=\Gamma(nm)\iL\Big[s^{-1-nm},\{s,t\}\Big]_{t=1}\int_{-\infty}^{\infty} x R_1(x)dx\\
&=\frac{1}{nm}\int_{-\infty}^{\infty} x R_1(x)dx.
\end{align}
In the second step above, we have used Eq.~\eqref{R1Laplace}. The third step follows via a simple scaling of $\lambda$ by $s$, and in the final step, we have used the inverse Laplace transform result $\iL[s^{-\gamma},\{s,t\}]_{t=1}=1/\Gamma(\gamma)$. The $x$-integral can now be performed by inserting $R_1(x)$ from Eq.~\eqref{R1x}; see Appendix~\ref{SecApp1}. We obtain,
\begin{equation}
\label{1st}
\left\langle\sum_{j=1}^{n}\lambda_{j}\right\rangle=\frac{1}{nm\,\Delta(\{a\})}~\sum_{i=1}^{n}\det[\zeta^{(i)}_{j,k}]_{j,k=1}^n,
\end{equation}
where
\begin{align}
\zeta^{(i)}_{j,k}=\begin{cases}a_{j}^{i-2}\,(m-i+1), & k=i,\\
 a_{j}^{k-1}, &  k\ne i.
\end{cases}
\end{align}

%%%%FIG1
\begin{figure*}[!t]
\advance\leftskip-1cm
\advance\rightskip-3cm
\includegraphics[width=1.0\linewidth]{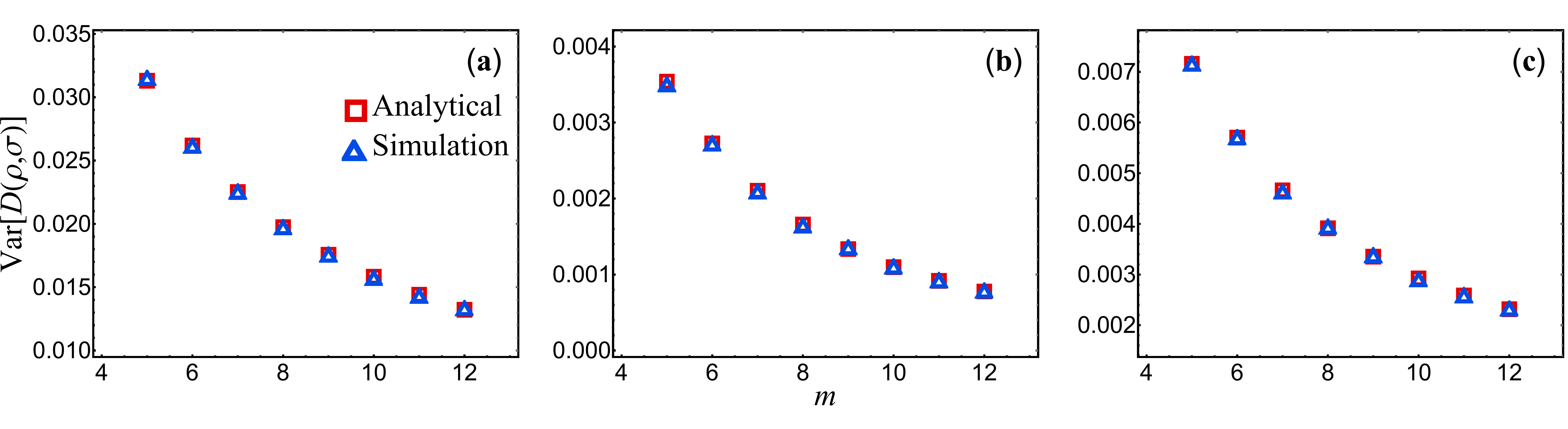}
\centering
\caption{Variance of the squared Bures distance between fixed density matrix $\sigma$ and random density matrix $\rho$ for various choices of $\sigma$: (a) a pure state, (b) a maximally mixed state, (c) a state with eigenvalues (0.08,0.10,0.12,0.28,0.42). The dimension of the two density matrices is set to be $n=5$ in all cases, and the dimension $m$ of the auxiliary subsystem associated with $\rho$ varies from $n$ to $n+7$.}
\label{varfix}
\end{figure*}
%%%%%
%%%%FIG2
\begin{figure*}[!t]
\advance\leftskip-1cm
\advance\rightskip-3cm
\includegraphics[width=1.0\linewidth]{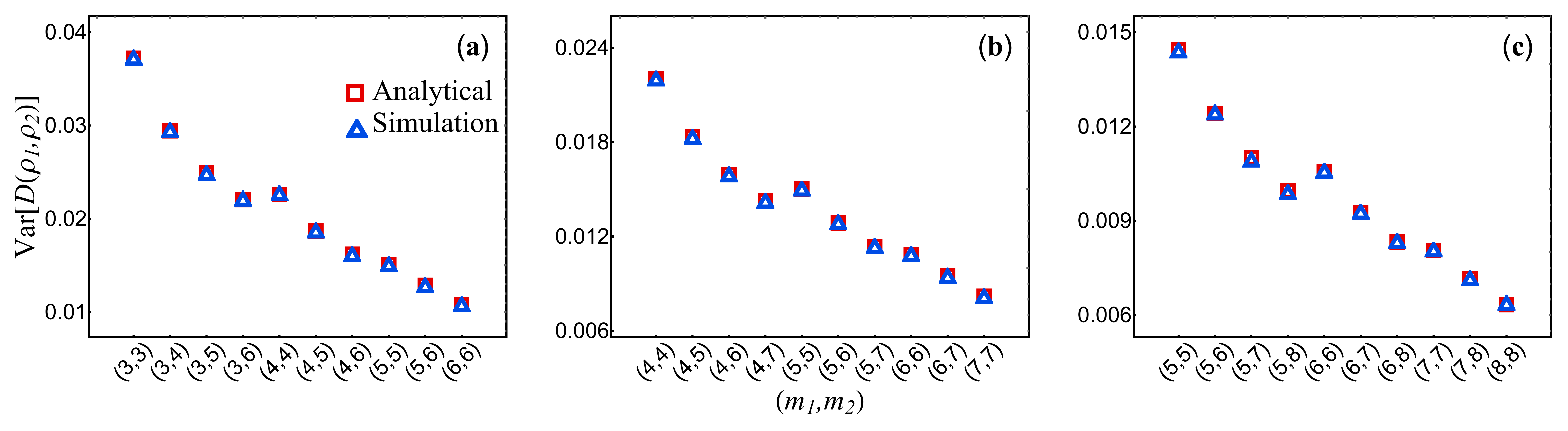}
\centering
\caption{Variance of the squared Bures distance between two independent random density matrices with (a) $n=3$,  (b) $n=4$ and (c) $n=5$ . In both cases various combinations of $m_1$ and $m_2$ values have been considered, as depicted on the horizontal axes.}
\label{varand}
\end{figure*}
%%%%%
%%% %

Our next task is to evaluate the second term in Eq.~\eqref{mF}, i.e., 
\begin{align}
\left\langle\sum_{{j,k=1}\atop{(j\ne k)}}^{n}\lambda_{j}^{1/2}\lambda_{k}^{1/2}\right\rangle=\int_{-\infty}^{\infty}\int_{-\infty}^{\infty}~\lambda_{1}^{1/2}\lambda_{2}^{1/2}~\cR_{2}(\lambda_1,\lambda_2)d\lambda_1d\lambda_2.
\end{align}
Following steps similar to those used in deriving Eq.~\eqref{R1avg}, we map the integrals over $\lambda_1,\lambda_2$ to those over $x_1,x_2$ and obtain,
\begin{align}
\label{lambdajkhalf}
\nonumber
&\left\langle\sum_{{j,k=1}\atop{(j\ne k)}}^{n}\lambda_{j}^{1/2}\lambda_{k}^{1/2}\right\rangle\\
\nonumber
&=\frac{1}{nm}\int_{-\infty}^{\infty}\int_{-\infty}^{\infty}~x_{1}^{1/2}x_{2}^{1/2}~R_{2}(x_1,x_2)dx_1dx_2\\
\nonumber
&=\frac{1}{nm}\int_{-\infty}^{\infty}~x_{1}^{1/2}~R_{1}(x_1)dx_1\int_{-\infty}^{\infty}~x_{2}^{1/2}~R_{1}(x_2)dx_2\\
&~~~ -\frac{1}{nm}\int_{-\infty}^{\infty}~x_{1}^{1/2}x_{2}^{1/2}~S(x_1,x_2)S(x_2,x_1)dx_1dx_2.
\end{align}
These integrals have also been performed in Appendix~\ref{SecApp1}. We get the final expression as,
\begin{align}
\label{2nd}
\nonumber
&\left\langle\sum_{{j,k=1}\atop{(j\ne k)}}^{n}\lambda_{j}^{1/2}\lambda_{k}^{1/2}\right\rangle=\frac{2}{nm\,\Delta^2(\{a\})}~~~~~~~\\
\nonumber
&\times\sum_{1\le i<l\le n}\Big(\det[\eta^{i,i}_{j,k}]_{j,k=1}^{n}\det[\eta^{l,l}_{j,k}]_{j,k=1}^{n}\\
&~~~~~~ -\det[\eta^{i,l}_{j,k}]_{j,k=1}^{n} \det[\eta^{l,i}_{j,k}]_{j,k=1}^{n}\Big),
\end{align}
where $\eta^{i,l}_{j,k}$ is defined as the following,
\begin{align}
\label{eta}
\eta^{i,l}_{j,k}=\begin{cases}a_j^{l-3/2}(m-l+1)_{1/2}, & k=i,\\
a_j^{k-1}, & k\neq i.\end{cases}
\end{align}
Here $(c)_d=\Gamma(c+d)/\Gamma(c)$ is the Pochhammer symbol.
Thus, overall the expression of $\expval{\cF}$ follows by using Eqs.~\eqref{1st} and~\eqref{2nd} in~\eqref{mF}. Next, we need the expression for the mean root fidelity, i.e.,
\begin{equation}
\langle\sqrt{\cF}\,\rangle=\expval{\sum_j \lambda_j^{1/2}}.
\end{equation}
It can be calculated similar to $\expval{\sum_j \lambda_j}$ using $\cR_1(\lambda)$, as has been already done in Ref.~\cite{LAK2021}. It is given by,
\begin{align}
\label{mrf1}
\langle\sqrt{\mathcal{F}}\,\rangle&=\frac{1}{(nm)_{1/2}\,\Delta(\{a\})}\sum_{i=1}^{n}\det[\eta^{i,i}_{j,k}]_{j,k=1}^n.
\end{align}
Finally, the variance of the squared Bures distance can be calcuated from Eq.~\eqref{var} by combining the expressions for $\langle\cF\rangle$ and $\langle\sqrt{\cF}\,\rangle$. One may encounter a situation where there are multiplicities in the eigenvalues of the fixed density matrix $\sigma$. This will result in a $0/0$ form in the expressions for the averages. In these cases, the analytical results can be obtained from Eqs.~\eqref{1st},\eqref{2nd} by invoking a limiting procedure. However, as far as numerical evaluation is concerned, one may use very close but unequal values for these degenerate eigenvalues.

Two cases of special interest are when the fixed density matrix $\sigma$ represents either a pure state or a maximally mixed state.  For pure state only one eigenvalue of the matrix $\sigma$ is equal to $1$ and the rest are 0. On the other hand, for the maximally mixed state all eigenvalues of $\sigma$ are $1/n$ and hence $\sigma=n^{-1}\mathds{1}_n$. In these cases the mean fidelity can be evaluated from Eqs~\eqref{1st},\eqref{2nd} by using limiting procedure, as indicated above. However, we may derive these results directly also, as done below.
%%%%FIG3
\begin{figure*}[!ht]
\advance\leftskip-1cm
\advance\rightskip-3cm
\includegraphics[width=1.0\linewidth]{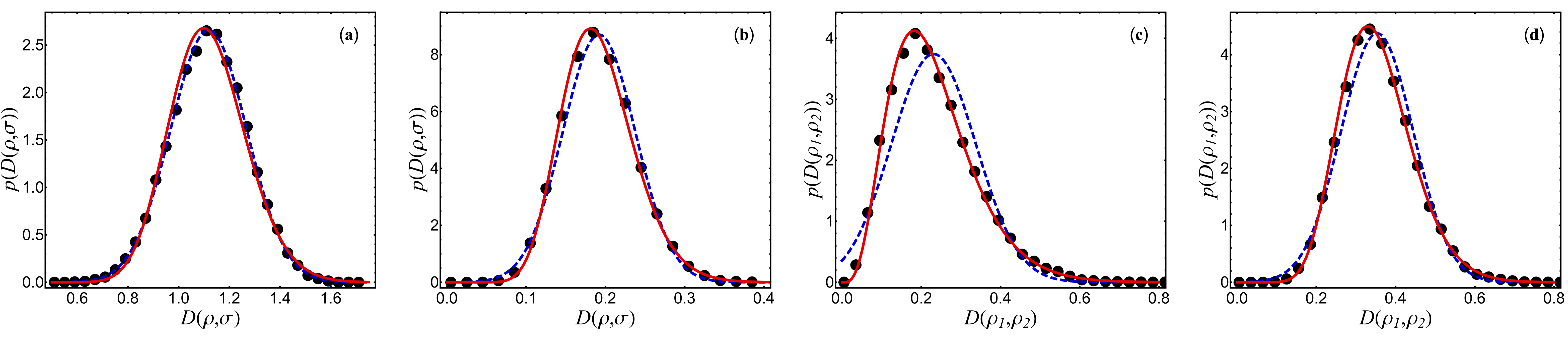}
\centering
\caption{Comparison of approximate PDF and the PDF obtained from Monte Carlo simulations for the squared Bures distance between (a) a random density matrix $\rho$ and a pure state $\sigma$ with $(n, m)=(5, 7)$; (b)a random density matrix $\rho$ and a maximally mixed state $\sigma$ with $(n, m)=(5, 7)$; (c), (d) two random density matrices $\rho_1$ and $\rho_2$ with $(n, m_1, m_2)=(3, 5, 7)$ and $(5, 6, 8)$, respectively. The solid lines represent gamma distribution based approximation and the dashed lines are for Gaussian distribution based approximation. The symbols (black disks) represent the PDFs obtained from Monte Carlo simulations.}
\label{den1}
\end{figure*}
%%%%%

\noindent
\emph{Pure state $\sigma$}:
In this case, only one eigenvalue of $\sigma$ is equal to $1$ and hence the only non-zero eigenvalue of the matrix $\sqrt{\sigma}\rho\sqrt{\sigma}$ coincides with the fidelity, whose PDF turns out to be~\cite{ZS2005,LAK2021}, 
\begin{align}
\label{pFpure}
p_\mathcal{F}(\mathcal{F})=\frac{\Gamma(n m)\mathcal{F}^{m-1}(1-\mathcal{F})^{nm-m-1}}{\Gamma(m)\Gamma(nm-m)}\Theta(\mathcal{F})\Theta(1-\mathcal{F}).
\end{align}
Therefore, the mean fidelity can be readily calculated as,
\begin{align}
\langle\mathcal{F}\rangle=\int_{-\infty}^{\infty}\mathcal{F}~p_{\mathcal{F}}(\mathcal{F})~d\mathcal{F}=\frac{1}{n}.
\end{align}
Similarly, the mean root fidelity is given by,
\begin{align}
\langle\mathcal{\sqrt{F}}\rangle=\int_{-\infty}^{\infty}\mathcal{F}^{1/2}~p_{\mathcal{F}}(\mathcal{F})~d\mathcal{F}=\frac{(m)_{1/2}}{(nm)_{1/2}}.
\end{align}
The variance of squared Bures distance then follows using Eq.~\eqref{var} and the above two expressions. In fact, in this case, the $\alpha$th moment of $\cF$ can be easily calculated to give $\langle\mathcal{F}^\alpha\rangle=(m)_{\alpha}/(nm)_{\alpha}$.

~\\
\noindent
\emph{Maximally mixed $\sigma$}:
In this case, the fixed density matrix is given by $\sigma=n^{-1}\mathds{1}_{n}$. Consequently, the eigenvalues of the matrix $\sqrt{\sigma}W\sqrt{\sigma}$ are $n^{-1}$ times the eigenvalues of the Wishart-Laguerre random matrix $W$. For the latter, the correlation functions are expressible in terms of associated Laguerre polynomials and that makes it possible to obtain in the present case an expression of the $\text{Var}(D)$ which is not determinant based. As shown in Appendix~\ref{SecApp2}, we find that the mean fidelity can be expressed as
%\begin{align}
%\nonumber
%&\langle\cF\rangle=\frac{1}{n}+\frac{1}{n m}[(nm)_{1/2}\,\langle\sqrt{\cF}\rangle]^2\\
%\nonumber
%&-\frac{\Gamma^2(m-n+3/2)}{n^2m}\sum_{j,k=0}^{n-1}\frac{j! \, k!}{(m-n+j)!(m-n+k)!}\\
%&\times\left[\binom{\frac{1}{2}}{j}\binom{\frac{1}{2}}{k} \,_3F_2\left(m-n+\tfrac{3}{2},-j,-k; \tfrac{3}{2}-j,\tfrac{3}{2}-k;1\right)\right]^2,
%\end{align}

\begin{align}
\label{mmixed}
\langle\cF\rangle=\frac{1}{n}+\frac{2\,\Gamma^2(m-n+\tfrac{3}{2})}{n^2m}\sum_{0\le j<k\le n-1}(\xi_{j,j}\xi_{k,k}-\xi_{j,k}\xi_{k,j}),
\end{align}
where
\begin{align}
\nonumber
\xi_{j,k}&=\frac{j! }{(m-n+j)!}\binom{\frac{1}{2}}{j}^2\\
&\times \,_3F_2\left(m-n+\tfrac{3}{2},-j,-k; \tfrac{3}{2}-j,\tfrac{3}{2}-k;1\right),
\end{align}
with $_3F_2(\cdots)$ being generalized Hypergeometric function. It may be noted that $\xi_{j,k}=\xi_{k,j}$. The mean root fidelity is given by~\cite{LAK2021},
\begin{align}
\nonumber
&\langle\sqrt{\cF}\rangle=\frac{1}{\sqrt{n}\,(nm)_{1/2}}\Gamma(m-n+\tfrac{3}{2})\sum_{j=0}^{n-1} \xi_{j,j}\\
&=\frac{2}{\sqrt{n}\,(nm)_{1/2}}\sum_{j=0}^{n-1}\binom{\frac{1}{2}}{j}\binom{\frac{1}{2}}{j+1}\frac{(m)_{1/2-j}}{(n+1)_{-j-1}}.
\end{align}
The variance of the squared Bures distance can now be evaluated using Eq.~\eqref{var}. 

In Fig.~\ref{varfix} we corroborate our analytical results for the variance of the squared Bures distance between a fixed density matrix $\sigma$ and a random density matrix $\rho$ by comparing them with Monte Carlo based numerical simulations.  We use three different choices of $\sigma$, which include the cases of pure state and maximally mixed state. In each case, we observe very good match between the analytical and simulation results.

%%% %
%%%% TABLE 1
\begin{table*}
\caption{Mean and variance of squared Bures distance between a pure state and a random state: Comparison between RMT analytical and coupled kicked top (CKT) simulation results using absolute differences $|\expval{D}_{CKT}-\expval{D}_{RMT}|$, $|\mathrm{Var}(D)_{CKT}-\mathrm{Var}(D)_{RMT}|$ and percent relative differences $|\expval{D}_{CKT}/\expval{D}_{RMT}-1|\times 100\%$, $|\mathrm{Var}(D)_{CKT}/\mathrm{Var}(D)_{RMT}-1|\times 100\%$. For each combination of $(n,m)$ values, we have considered three sets of stochasticity and coupling parameters ($\kappa_1,\kappa_2,\epsilon$) for the coupled kicked tops, viz., CKT 1: (6,7,0.75), CKT 2: (7,8,1), CKT 3: (7,9,0.5).}
\begin{tabular}{|c|c|c|c|c|c|c|c|c|c|}
\hline
\multirow{2}{*}{$n$} & \multirow{2}{*}{$m$} & \multicolumn{4}{c|}{$\langle D \rangle $} & %
    \multicolumn{4}{c|}{Var$(D)$} \\
\cline{3-10}
& & RMT & CKT & Abs. Diff. & Rel. Diff.(\%) & RMT & CKT & Abs. Diff. & Rel. Diff.(\%) \\
\hline
\multirow{3}{*}{$21$} & \multirow{3}{*}{$21$} & \multirow{3}{*}{$1.56603$} & 1.56832 & $2.29\times 10^{-3}$ & $1.46\times 10^{-1}$& \multirow{3}{*}{$2.15\times 10^{-3}$} & $2.09\times 10^{-3}$ & $6.00\times 10^{-5}$ & $2.79$ \\
\cline{4-6}\cline{8-10}
~ & ~& ~& 1.57326 &  $7.23\times 10^{-3}$& $4.62\times 10^{-1}$ & {} & $2.04\times 10^{-3}$ & $1.10\times 10^{-4}$ & 5.11 \\
\cline{4-6}\cline{8-10}
~ & ~& ~& 1.57398 &  $7.95\times 10^{-3}$ & $5.08\times 10^{-1}$ & {} & $1.95\times 10^{-3}$ & $2.00\times 10^{-4}$ & 9.30 \\
\hline
\multirow{3}{*}{$21$} & \multirow{3}{*}{$31$} & \multirow{3}{*}{$1.56524$} & 1.57443 & $9.19\times 10^{-3}$ & $5.87\times 10^{-1}$& \multirow{3}{*}{$1.46\times 10^{-3}$} & $1.35\times 10^{-3}$ & $1.10\times 10^{-4}$ & 7.53  \\
\cline{4-6}\cline{8-10}
~ & ~& ~& 1.56312 &  $2.12\times 10^{-3}$ & $1.35\times 10^{-1}$ & {} & $1.45\times 10^{-3}$ & $1.00\times 10^{-5}$ & $6.84\times 10^{-1}$ \\
\cline{4-6}\cline{8-10}
~ & ~& ~& 1.57311 &  $7.87\times 10^{-3}$ & $5.03\times 10^{-1}$ & {} & $1.42\times 10^{-3}$ & $4.00\times 10^{-5}$ & 2.73\\
\hline
\multirow{3}{*}{$21$} & \multirow{3}{*}{$41$} & \multirow{3}{*}{$1.56483$} & 1.57436 & $9.53\times 10^{-3}$ & $6.09\times 10^{-1}$& \multirow{3}{*}{$1.10\times 10^{-3}$} & $1.04\times 10^{-3}$ & $6.00\times 10^{-5}$ & 5.45  \\
\cline{4-6}\cline{8-10}
~ & ~& ~& 1.56484 &  $1.00\times 10^{-5}$ & $6.39\times 10^{-4}$ & {} & $1.09\times 10^{-3}$ & $1.00\times 10^{-5}$ & $9.09\times 10^{-1}$ \\
\cline{4-6}\cline{8-10}
~ & ~& ~& 1.56701 &  $2.18\times 10^{-3}$ & $1.39\times 10^{-1}$ & {} & $1.12\times 10^{-3}$ & $2.00\times 10^{-5}$ & 1.81\\
\hline
\multirow{3}{*}{$23$} & \multirow{3}{*}{$23$} & \multirow{3}{*}{$1.58513$} & 1.59099 & $5.86\times 10^{-3}$ & $3.69\times 10^{-1}$& \multirow{3}{*}{$1.79\times 10^{-3}$} & $1.72\times 10^{-3}$ & $7.00\times 10^{-5}$ & 3.91  \\
\cline{4-6}\cline{8-10}
~ & ~& ~& 1.59460 &  $9.47\times 10^{-3}$ & $5.97\times 10^{-1}$ & {} & $1.69\times 10^{-3}$ & $1.00\times 10^{-4}$ & 5.58 \\
\cline{4-6}\cline{8-10}
~ & ~& ~& 1.59480 &  $9.67\times 10^{-3}$ & $6.10\times 10^{-1}$ & {} & $1.70\times 10^{-3}$ & $9.00\times 10^{-5}$ & 5.14 \\
\hline
\multirow{3}{*}{$23$} & \multirow{3}{*}{$33$} & \multirow{3}{*}{$1.58448$} & 1.58444 & $4.00\times 10^{-5}$ & $2.52\times 10^{-3}$& \multirow{3}{*}{$1.25\times 10^{-3}$} & $1.23\times 10^{-3}$ & $2.00\times 10^{-5}$ & 1.60  \\
\cline{4-6}\cline{8-10}
~ & ~& ~& 1.58999 &  $5.51\times 10^{-3}$ & $3.48\times 10^{-1}$ & {} & $1.22\times 10^{-3}$ & $3.00\times 10^{-5}$ & 2.40 \\
\cline{4-6}\cline{8-10}
~ & ~& ~& 1.59078 &  $6.30\times 10^{-3}$ & $3.97\times 10^{-1}$ & {} & $1.21\times 10^{-3}$ & $4.00\times 10^{-5}$ & 3.20 \\
\hline
\multirow{3}{*}{$23$} & \multirow{3}{*}{$43$} & \multirow{3}{*}{$1.58413$} & 1.59036 & $6.23\times 10^{-3}$ & $3.93\times 10^{-1}$& \multirow{3}{*}{$9.64\times 10^{-4}$} & $9.34\times 10^{-4}$ & $3.00\times 10^{-5}$ & 3.11  \\
\cline{4-6}\cline{8-10}
~ & ~& ~& 1.58776 &  $3.63\times 10^{-3}$ & $2.29\times 10^{-1}$ & {} & $9.48\times 10^{-4}$ & $1.60\times 10^{-5}$ & 1.65 \\
\cline{4-6}\cline{8-10}
~ & ~& ~& 1.58332 &  $8.10\times 10^{-4}$ & $5.11\times 10^{-2}$ & {} & $9.68\times 10^{-4}$ & $4.00\times 10^{-6}$ & $4.14\times 10^{-1}$ \\

\hline
\end{tabular}
\label{CKTps}
\end{table*}

%%% %
%%%% TABLE 2
\begin{table*}
\caption{Mean and variance of squared Bures distance between a maximally mixed state and a random state: Comparison between RMT and coupled kicked top results using absolute differences and percent relative differences. Here also we have considered three sets of stochasticity and coupling parameters as mentioned in the caption of TABLE~\ref{CKTps}.}
\begin{tabular}{|c|c|c|c|c|c|c|c|c|c|}
\hline
\multirow{2}{*}{$n$} & \multirow{2}{*}{$m$} & \multicolumn{4}{c|}{$\langle D \rangle $} & %
    \multicolumn{4}{c|}{Var$(D)$} \\
\cline{3-10}
& & RMT & CKT & Abs. Diff. & Rel. Diff.(\%) & RMT & CKT & Abs. Diff. & Rel. Diff.(\%) \\
\hline
\multirow{3}{*}{$21$} & \multirow{3}{*}{$21$} & \multirow{3}{*}{$3.01\times 10^{-1}$} & $3.03\times 10^{-1}$ & $2.00\times 10^{-3}$ & $6.64\times 10^{-1}$& \multirow{3}{*}{$2.05\times 10^{-4}$} & $2.15\times 10^{-4}$ & $1.00\times 10^{-5}$ & 4.87  \\
\cline{4-6}\cline{8-10}
~ & ~& ~& $3.02\times 10^{-1}$&  $1.00\times 10^{-3}$ & $3.32\times 10^{-1}$ & {} & $2.09\times 10^{-4}$ & $4.00\times 10^{-6} $& 1.95 \\
\cline{4-6}\cline{8-10}
~ & ~& ~& $3.04\times 10^{-1}$ &  $3.00\times 10^{-3}$ & $9.96\times 10^{-1}$ & {} & $2.11\times 10^{-4}$ & $6.00\times 10^{-6}$ & 2.92 \\
\hline
\multirow{3}{*}{$21$} & \multirow{3}{*}{$23$} & \multirow{3}{*}{$2.67\times 10^{-1}$} & $2.69\times 10^{-1}$ & $2.00\times 10^{-3}$ & $7.49\times 10^{-1}$& \multirow{3}{*}{$1.84\times 10^{-4}$} & $1.96\times 10^{-4}$ & $1.20\times 10^{-5}$ & 6.52  \\
\cline{4-6}\cline{8-10}
~ & ~& ~& $2.68\times 10^{-1}$ &  $1.00\times 10^{-3}$ & $3.74\times  10^{-1}$ & {} & $1.85\times 10^{-4}$ & $1.00\times 10^{-6}$ & $5.43\times 10^{-1}$ \\
\cline{4-6}\cline{8-10}
~ & ~& ~& $2.71\times 10^{-1}$ &  $4.00\times 10^{-3}$ & $1.49$ & {} & $1.91\times 10^{-4}$ & $7.00\times 10^{-6}$ & 3.80 \\
\hline
\multirow{3}{*}{$21$} & \multirow{3}{*}{$25$} & \multirow{3}{*}{$2.41\times 10^{-1}$} & $2.43\times 10^{-1}$ & $2.00\times 10^{-3}$ & $8.29\times 10^{-1}$& \multirow{3}{*}{$1.63\times 10^{-4}$} & $1.73\times 10^{-4}$& $1.00\times 10^{-5}$ & 6.13  \\
\cline{4-6}\cline{8-10}
~ & ~& ~& $2.42\times 10^{-1}$ &  $1.00\times 10^{-3}$ & $4.14\times 10^{-1}$ & {} & $1.65\times 10^{-4}$ & $2.00\times 10^{-6}$ & 1.23 \\
\cline{4-6}\cline{8-10}
~ & ~& ~& $2.44\times 10^{-1}$ &  $3.00\times 10^{-3}$ & $1.24$ & {} & $1.68\times 10^{-4}$ & $5.00\times 10^{-6}$ & $3.06$ \\
\hline
\multirow{3}{*}{$23$} & \multirow{3}{*}{$23$} & \multirow{3}{*}{$3.02\times 10^{-1}$} & $3.01\times 10^{-1}$& $1.00\times 10^{-3}$ & $3.31\times 10^{-1}$& \multirow{3}{*}{$1.71\times 10^{-4}$} & $1.86\times 10^{-4}$ & $1.50\times 10^{-5}$ & 8.77  \\
\cline{4-6}\cline{8-10}
~ & ~& ~& $3.05\times 10^{-1}$ &  $3.00\times 10^{-3}$ & $9.93\times 10^{-1}$ & {} & $1.72\times 10^{-4}$ & $1.00\times 10^{-6}$ & $5.84\times 10^{-1}$ \\
\cline{4-6}\cline{8-10}
~ & ~& ~& $3.04\times 10^{-1}$ &  $2.00\times 10^{-3}$ & $6.62\times 10^{-1}$ & {} & $1.75\times 10^{-4}$ & $4.00\times 10^{-6}$ & $2.34$ \\
\hline
\multirow{3}{*}{$23$} & \multirow{3}{*}{$25$} & \multirow{3}{*}{$2.70\times 10^{-1}$} & $2.72\times 10^{-1}$ & $2.00\times 10^{-3}$ & $7.40\times 10^{-1}$& \multirow{3}{*}{$1.55\times 10^{-4}$} & $1.42\times 10^{-4}$ & $1.30\times 10^{-5}$ & 8.38 \\
\cline{4-6}\cline{8-10}
~ & ~& ~& $2.71\times 10^{-1}$ & $1.00\times 10^{-3}$ & $3.70\times 10^{-1}$ & {} & $1.53\times 10^{-4}$ & $2.00\times 10^{-6}$ & 1.29 \\
\cline{4-6}\cline{8-10}
~ & ~& ~& $2.71\times 10^{-1}$& $1.00\times 10^{-3}$ & $3.70\times 10^{-1}$ & {} & $1.56\times 10^{-4}$ & $1.00\times 10^{-6}$ & $6.45\times 10^{-1}$ \\
\hline
\multirow{3}{*}{$23$} & \multirow{3}{*}{$27$} & \multirow{3}{*}{$2.45\times 10^{-1}$} & $2.47\times 10^{-1}$ &$2.00\times 10^{-3}$ & $8.16\times 10^{-1}$& \multirow{3}{*}{$1.39\times 10^{-4}$} & $1.49\times 10^{-4}$ & $1.00\times 10^{-5}$ & 7.19  \\
\cline{4-6}\cline{8-10}
~ & ~& ~& $2.46\times 10^{-1}$ & $1.00\times 10^{-3}$ & $4.08\times 10^{-1}$ & {} & $1.38\times 10^{-4}$ & $1.00\times 10^{-6}$ & $7.19\times 10^{-1}$ \\
\cline{4-6}\cline{8-10}
~ & ~& ~& $2.47\times 10^{-1}$ &  $2.00\times 10^{-3}$ & $8.16\times 10^{-1}$ & {} & $1.41\times 10^{-4}$ & $2.00\times 10^{-6}$ & 1.44 \\
\hline
\end{tabular}
\label{CKTmms}
\end{table*}
%%% %
%%%% TABLE 3
\begin{table*}
\caption{Mean and variance of squared Bures distance between two random states: Comparison between RMT and coupled kicked top pair (CKTP) results.  For each combinations of ($n,m_1,m_2$) values, the CKTP simulations have been carried out for three sets of parameters ($\kappa_{1}^{A}, \kappa_{2}^{A}, \epsilon^{A};\kappa_{1}^{B}, \kappa_{2}^{B}, \epsilon^{B}$), viz., CKTP 1: ($8,7,0.5;7,8,1$), CKTP 2: ($6,7,0.8;6,8,0.9$), CKTP 3: ($7,8,0.75;8,7,0.85$).}
\begin{tabular}{|c|c|c|c|c|c|c|c|c|c|c|}
\hline
\multirow{2}{*}{$n$} & \multirow{2}{*}{$m_1$} & \multirow{2}{*}{$m_2$} & \multicolumn{4}{c|}{$\langle D \rangle $} & %
    \multicolumn{4}{c|}{Var$(D)$} \\
\cline{4-11}
& &  & RMT & CKTP & Abs. Diff. & Rel. Diff.(\%) & RMT & CKPT & Abs. Diff. & Rel. Diff.(\%) \\
\hline
\multirow{3}{*}{$21$} & \multirow{3}{*}{$21$} & \multirow{3}{*}{$21$} & \multirow{3}{*}{$4.989\times 10^{-1}$} & $4.994\times 10^{-1}$ & $5.00\times 10^{-4}$ & $1.00\times 10^{-1}$& \multirow{3}{*}{$8.48\times 10^{-4}$} & $8.62\times 10^{-4}$ & $1.40\times 10^{-5}$ & 1.65  \\
\cline{5-7}\cline{9-11}
~ &~ & ~& ~& $4.993\times 10^{-1}$ &  $4.00\times 10^{-4}$ & $8.01\times 10^{-2}$ & {} & $8.43\times 10^{-4}$ & $5.00\times 10^{-6}$ & $5.89\times 10^{-1}$ \\
\cline{5-7}\cline{9-11}
~ & ~ & ~& ~& $4.998\times 10^{-1}$ &  $9.00\times 10^{-4}$ & $1.80\times 10^{-1}$ & {} & $8.47\times 10^{-4}$ & $1.00\times 10^{-6}$ & $1.17\times 10^{-1}$ \\
\hline
\multirow{3}{*}{$21$} & \multirow{3}{*}{$21$} & \multirow{3}{*}{$25$} & \multirow{3}{*}{$4.615\times 10^{-1}$} & $4.622\times 10^{-1}$& $7.00\times 10^{-4}$ & $1.51\times 10^{-1}$& \multirow{3}{*}{$7.34\times 10^{-4}$} & $7.40\times 10^{-4}$ & $6.00\times 10^{-6}$ & $8.17\times 10^{-1}$  \\
\cline{5-7}\cline{9-11}
~ &~ & ~& ~& $4.616\times 10^{-1}$ &  $1.00\times 10^{-4}$ & $2.16\times 10^{-2}$ & {} & $7.33\times 10^{-4}$ & $1.00\times 10^{-6}$ & $1.36\times 10^{-1}$\\
\cline{5-7}\cline{9-11}
~ & ~ & ~& ~& $4.626\times 10^{-1}$ & $1.10\times 10^{-3}$ & $2.38\times 10^{-1}$ & {} & $7.48\times 10^{-4}$ & $1.40\times 10^{-5}$ & $1.90$ \\
\hline
\multirow{3}{*}{$21$} & \multirow{3}{*}{$23$} & \multirow{3}{*}{$25$} & \multirow{3}{*}{$4.378\times 10^{-1}$} & $4.381\times 10^{-1}$ & $3.00\times 10^{-4}$ & $6.85\times 10^{-2}$& \multirow{3}{*}{$6.78\times 10^{-4}$} & $6.90\times10^{-4}$ & $1.20\times 10^{-5}$ & $1.77$  \\
\cline{5-7}\cline{9-11}
~ &~ & ~& ~& $4.375\times 10^{-1}$ & $3.00\times 10^{-4}$ & $6.85\times 10^{-2}$ & {} & $6.73\times 10^{-4}$ & $5.00\times 10^{-6}$ & $7.37\times 10^{-1}$ \\
\cline{5-7}\cline{9-11}
~ & ~ & ~& ~& $4.385\times 10^{-1}$ & $7.00\times 10^{-4}$ & $1.59\times 10^{-1}$ & {} & $6.70\times 10^{-4}$ & $8.00\times 10^{-6}$ & 1.18 \\
\hline
\multirow{3}{*}{$23$} & \multirow{3}{*}{$23$} & \multirow{3}{*}{$23$} & \multirow{3}{*}{$4.991\times 10^{-1}$} & $4.993\times 10^{-1}$& $2.00\times 10^{-4}$ & $4.00\times 10^{-2}$& \multirow{3}{*}{$7.07\times 10^{-4}$} & $7.14\times 10^{-4}$ & $7.00\times 10^{-6}$ & $9.90\times 10^{-1}$  \\
\cline{5-7}\cline{9-11}
~ &~ & ~& ~& $4.996\times 10^{-1}$ & $5.00\times 10^{-4}$ & $1.00\times 10^{-1}$ & {} & $7.09\times 10^{-4}$ & $2.00\times 10^{-6}$ & $2.82\times 10^{-1}$ \\
\cline{5-7}\cline{9-11}
~ & ~ & ~& ~& $5.000\times 10^{-1}$ & $9.00\times 10^{-4}$ & $1.80\times 10^{-1}$ & {} & $7.04\times 10^{-4}$ & $3.00\times 10^{-6}$ & $4.24\times 10^{-1}$ \\
\hline
\multirow{3}{*}{$23$} & \multirow{3}{*}{$23$} & \multirow{3}{*}{$25$} & \multirow{3}{*}{$4.798\times 10^{-1}$} & $4.803\times 10^{-1}$ & $5.00\times 10^{-4}$ & $1.04\times 10^{-1}$& \multirow{3}{*}{$6.60\times 10^{-4}$} & $6.59\times 10^{-4}$ & $1.00\times 10^{-6}$ & $1.51\times 10^{-1}$  \\
\cline{5-7}\cline{9-11}
~ &~ & ~& ~& $4.799\times 10^{-1}$ & $1.00\times 10^{-4}$ & $2.08\times 10^{-2}$ & {} & $6.62\times 10^{-4}$ & $2.00\times 10^{-6}$ & $3.03\times 10^{-1}$ \\
\cline{5-7}\cline{9-11}
~ & ~ & ~& ~& $4.804\times 10^{-1}$ & $6.00\times 10^{-4}$ & $1.25\times 10^{-1}$ & {} & $6.68\times 10^{-4}$ & $8.00\times 10^{-6}$ & 1.21 \\
\hline
\multirow{3}{*}{$23$} & \multirow{3}{*}{$25$} & \multirow{3}{*}{$27$} & \multirow{3}{*}{$4.426\times 10^{-1}$} & $4.427\times 10^{-1}$ & $1.00\times 10^{-4}$ & $2.25\times 10^{-2}$& \multirow{3}{*}{$5.76\times 10^{-4}$} & $5.74\times 10^{-4}$ & $2.00\times 10^{-6}$ & $3.47\times 10^{-1}$  \\
\cline{5-7}\cline{9-11}
~ &~ & ~& ~& $4.431\times 10^{-1}$ & $5.00\times 10^{-4}$ & $1.12\times 10^{-1}$ & {} & $5.67\times 10^{-4}$ & $9.00\times 10^{-6}$ & $1.56$ \\
\cline{5-7}\cline{9-11}
~ & ~ & ~& ~& $4.421\times 10^{-1}$ & $5.00\times 10^{-4}$ & $1.12\times 10^{-1}$ & {} & $5.75\times 10^{-4}$ & $1.00\times 10^{-6}$ & $1.73\times 10^{-1}$ \\
\hline
\end{tabular}
\label{CKT2r}
\end{table*}
%%% %

%%%%%%
\subsection{Two random density matrices $\rho_1$ and $\rho_2$}
%%%%%%
We now consider two independent random density matrices, $\rho_1$ and $\rho_2$, taken from the distribution given in Eq.~\eqref{dist} but with different $m$ values in general, say $m_1$ and $m_2$. To evaluate the desired quantities in this case, we need the information about the eigenvalue statistics of the random matrix $\sqrt{\rho_1}\rho_2 \sqrt{\rho_1}$ --- the one-point and two-point correlation functions to be precise. Again, these are related to the corresponding correlation functions of the random matrix $\sqrt{W_1}W_2 \sqrt{W_1}$ which involves two independent Wishart matrices $W_1$ and $W_2$. The eigenvalues of $\sqrt{W_1}W_2 \sqrt{W_1}$ are identical to the product matrix $W_1W_2$ or $W_2W_1$. Considering the unordered eigenvalues $\{y_i\}$ of $\sqrt{W_1}W_2 \sqrt{W_1}$, the first two correlation functions are given by,
\begin{align}
\label{Rt1y}
&\tR_1(y)=\tS(y,y),\\
\label{Rt2y}
&\tR_2(y_1,y_2)=\tR_1(y_1)\tR_1(y_2)-\tS(y_1,y_2)\tS(y_2,y_1).
\end{align}
In the above expressions, the kernel $\tS(y_1,y_2)$ is known in terms of Meijer $G$-functions as~\cite{AIK2013},
\begin{align}
\label{Sy1y2}
\nonumber
\tS(y_1,y_2)&=\sum_{j=0}^{n-1}G_{1,3}^{1,0} 
\left(
\begin{matrix}
j+1\\
0;-v_{1},-v_{2}
\end{matrix} \bigg| y_1\right)\\
&\times
G_{1,3}^{2,1} 
\left(
\begin{matrix}
-j\\
v_{1},v_{2};0
\end{matrix} \bigg| y_2\right)\Theta(y_1)\Theta(y_2)\\
\nonumber
&=\sum_{j=0}^{n-1}\sum_{k=0}^j\frac{(-y_1)^k}{k!(k+v_1)!(k+v_2)!(j-k)!}\\
\label{MGint}
&\times G_{1,3}^{2,1} 
\left(
\begin{matrix}
-j\\
v_{1},v_{2};0
\end{matrix} \bigg| y_2\right)\Theta(y_1)\Theta(y_2),
\end{align}
where $v_1=m_1-n$ and $v_2=m_2-n$. The second expression in the above equation follows using the expansion of $G_{1,3}^{1,0} (\cdots)$~\cite{AIK2013}. The corresponding correlation functions for the unordered eigenvalues $\{\mu_i\}$ of $\sqrt{\rho_1}\rho_2 \sqrt{\rho_1}$ are related to the above via a dual Laplace inversion~\cite{LAK2021},
\begin{align}
\label{cor1}
\nonumber
\tcR_1(\mu)=\Gamma(nm_1)\Gamma(nm_2)\iL\Big[s_1^{1-nm_1}s_2^{1-nm_2}\\
\times \tR_{1}(s_1s_2\mu), \{s_1,t_1\},\{s_2,t_2\}\Big]_{t_1=t_2=1},
\end{align}
\begin{align}
\label{cor2}
\nonumber
\tcR_2(\mu_1,\mu_2)=\Gamma(nm_1)\Gamma(nm_2)\mathcal{L}^{-1}\Big[s_1^{2-nm_1}s_2^{2-nm_2}\\
\times \widetilde{R}_{2}(s_1s_2\mu_1,s_1s_2\mu_2), \{s_1,t_1\},\{s_2,t_2\}\Big]_{t_1=t_2=1}.
\end{align}

Similar to Eq.~\eqref{mF}, in the present case, the mean fidelity can be evaluated as
\begin{align}
\label{mff}
\nonumber
\langle\mathcal{F}\rangle&=\left\langle\left(\tr \sqrt{\sqrt{\rho_1}\rho_2 \sqrt{\rho_1}}\right)^{2}\right\rangle
=\left\langle\sum_{j,k=1}^{n}\mu_{j}^{1/2}\mu_{k}^{1/2}\right\rangle\\
&=\left\langle\sum_{j=1}^{n}\mu_{j}\right\rangle+\left\langle\sum_{{j,k=1}\atop{(j\ne k)}}^{n}\mu_{j}^{1/2}\mu_{k}^{1/2}\right\rangle.
\end{align}
The first average in Eq.~\eqref{mff} can be calculated as,
\begin{align}
\label{mff1}
\nonumber
\left\langle\sum_{j=1}^{n}\mu_{j}\right\rangle&=\int_{-\infty}^{\infty}\mu \tcR_1(\mu)d\mu\\
&=\frac{1}{n^2m_1m_2} \int_{-\infty}^{\infty}y \tR_1(y)dy.
\end{align}
For the evaluation of the integral in this equation, we may use Eq.~\eqref{Rt1y}, which can be written in terms of Meijer $G$-function as given in Eq.~\eqref{Sy1y2}. However, it can be much easily evaluated using already known average over Wishart ensemble, viz., $\expval{\tr(XW)}_W=m\tr X$, where $X$ is Hermitian. Since $W_1$ and $W_2$ are independent, we have $\int_{-\infty}^\infty y \tR_1(y)dy=\expval{\tr(W_1^{1/2}W_2W_1^{1/2})}_{W_1,W_2}=\expval{\tr(W_1W_2)}_{W_1,W_2}=m_1\expval{\tr W_2}_{W_2}=m_1 m_2\tr \1_n=nm_1m_2$. Hence, we get a very simple result for Eq.~\eqref{mff1}, viz.,
\begin{align}
\label{mff1a}
\left\langle\sum_{j=1}^{n}\mu_{j}\right\rangle=\frac{1}{n}.
\end{align}

The second average in Eq.~\eqref{mff} can be written as
\begin{align}
\label{mff2}
\nonumber
&\left\langle\sum_{{j,k=1}\atop{(j\ne k)}}^{n}\mu_{j}^{1/2}\mu_{k}^{1/2}\right\rangle=\int_{-\infty}^\infty\int_{-\infty}^\infty \mu_1^{1/2}\mu_2^{1/2} \tcR_2(\mu_1,\mu_2)\\
\nonumber
&=\frac{1}{n^2m_1m_2}\int_{-\infty}^\infty \int_{-\infty}^\infty y_1^{1/2}y_2^{1/2}\tR_2(y_1,y_2)\\
\nonumber
&=\frac{1}{n^2m_1m_2}\int_{-\infty}^\infty y_1^{1/2}\tR_1(y_1)dy_1\int_{-\infty}^\infty y_2^{1/2}\tR_1(y_2)dy_2\\
& -\frac{1}{n^2m_1m_2}\int_{-\infty}^\infty \int_{-\infty}^\infty y_1^{1/2}y_2^{1/2}\tS(y_1,y_2)\tS(y_2,y_1)dy_1dy_2.
\end{align}
These integrals have been evaluated in Appendix~\ref{SecApp3}. On combining the ensuing expression with Eq.~\eqref{mff1a}, we obtain the mean fidelity in this case as
%\begin{widetext}
%\textcolor{blue}{
%\begin{align}
%\nonumber
%&\langle\mathcal{F}\rangle=\frac{1}{n}+\frac{1}{n^2m_1m_2}\left[(nm_1)_{1/2}(nm_2)_{1/2}\langle\sqrt{\cF}\rangle \right]^2\\
%\nonumber
%&-\frac{1}{n^2m_1m_2\pi^2}\sum_{j,k=1}^n\frac{1}{1/4-(j-k)^2}\\
%\nonumber
%&\times (j)_{1/2}(j+v_1)_{1/2}(j+v_2)_{1/2}(n-j+1)_{-1/2}\\
%&\times (k)_{1/2}(k+v_1)_{1/2}(k+v_2)_{1/2}(n-k+1)_{-1/2}.
%\end{align}}
%\end{widetext}
\begin{align}
\label{mf2}
\nonumber
&\langle\cF\rangle=\frac{1}{n}+\frac{8}{\pi^2 n^2m_1m_2}\sum_{1\le j<k\le n}\frac{(j-k)^2}{(j-k)^2-1/4}\\
\nonumber
&\times (j)_{1/2}(j+v_1)_{1/2}(j+v_2)_{1/2}(n-j+1)_{-1/2}\\
&\times (k)_{1/2}(k+v_1)_{1/2}(k+v_2)_{1/2}(n-k+1)_{-1/2}.
\end{align}

Next, the mean root fidelity is calculated as
\begin{align}
\label{mrf2}
\nonumber
&\langle\sqrt{\cF}\rangle=\left\langle \sum_{j=1}^n \mu_j^{1/2}\right\rangle=\int_{-\infty}^\infty \mu^{1/2}\tcR(\mu)d\mu\\
&=\frac{2}{\pi (n m_1)_{1/2}(n m_2)_{1/2}}\int_{-\infty}^\infty y^{1/2}\tR(y)dy.
\end{align}
As discussed in Appendix~\ref{SecApp3}, this gives
\begin{align}
\nonumber
&\langle\sqrt{\cF}\rangle=\frac{2}{\pi (n m_1)_{1/2}(n m_2)_{1/2}}\\
&\times\sum_{j=1}^n  (j)_{1/2}(j+v_1)_{1/2}(j+v_2)_{1/2}(n-j+1)_{-1/2}.
\end{align}
The above expression for the mean root fidelity is of slightly different form compared to that given in Ref.~\cite{LAK2021}. However, they are equivalent and the latter can be derived from the above expression using the identity $(n-j+1)_{-1/2}=\pi (-1)^{n-j}/[\Gamma(n-j+1)\Gamma(j-n+1/2)]$.

The variance of the squared Bures distance between two independent random density matrices can be obtained by using Eqs.~\eqref{mf2} and~\eqref{mrf2} in Eq.~\eqref{var}. In Fig.~\ref{varand} we contrast our analytical result with Monte Carlo simulations and find very good agreement.

%%%%FIG4
\begin{figure*}[!t]
\advance\leftskip-1cm
\advance\rightskip-3cm
\includegraphics[width=1.0\linewidth]{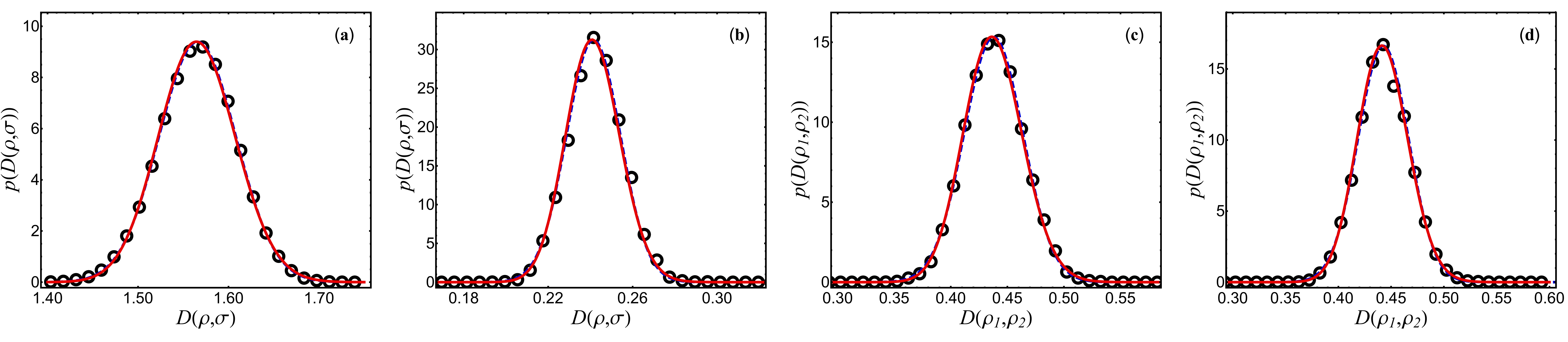}
\centering
\caption{Comparison of PDF approximation and the PDF obtained from coupled-kicked-tops simulations for the squared Bures distance between (a) a random density matrix $\rho$ and a pure state $\sigma$ with $(n, m)=(21, 25)$; (b) a random density matrix $\rho$ and a maximally mixed state $\sigma$ with $(n, m)=(21, 25)$; and (c), (d) two random density matrices $\rho_1$ and $\rho_2$ with $(n, m_1, m_2)=(21, 23, 25)$ and $(23, 25, 27)$ respectively. In the plots, solid lines represent gamma distribution based approximation and the dashed lines are for Gaussian distribution based approximation. The symbols (circles) represent the PDF obtained from coupled-kicked-tops simulations.}
\label{CKTpdf}
\end{figure*}
%%%%%

%%%%%
%%%%%%%%%
\section{Cumulants-based approximation of the PDF of squared Bures distance}
\label{sec3}
%%%%%%%%%

A random variable is completely characterized by its probability density function. The squared Bures distance $D$ between density matrices (one fixed one random, or both random), as considered in this work, is a random variable, so this also holds for it. In this section, we obtain approximate PDF for the same based on gamma distribution using the cumulants matching approach. The choice of gamma distribution for the approximation instead of Gaussian distribution stems from the fact that $D$ is a non-negative quantity.

We have,
\begin{align}
\label{gm}
p(D)=\frac{\gamma^{\kappa}}{\Gamma(\kappa)}D^{\kappa-1}e^{-\gamma D},
\end{align}
where $\kappa (>0)$ and $\gamma (>0)$ are the shape and rate parameters, respectively. The corresponding average and variance are given by $\kappa/\gamma$ and $\kappa/\gamma^2$ By matching these two with the average and variance of $D$ obtained earlier, we can determine the two parameters and thereby use the PDF in Eq.~\eqref{gm} as an approximation to the actual PDF of $D$. We have the relations,  
\begin{align}
\label{rel1}
\gamma=\frac{\langle D\rangle}{\mathrm{Var}(D)},~~~~
\kappa=\langle D\rangle \gamma.
\end{align}
In Fig.~\ref{den1} we compare the above gamma distribution-based approximation by comparing it with the PDF of $D$ obtained with the aid of Monte Carlo simulations for various cases. We find that the approximation works rather well. Out of curiosity, in the plots, we have also included the Gaussian distribution based approximation for the PDF of $D$ by plugging-in its mean and variance in the Gaussian function. Inevitably, the left tail of the Gaussian approximation does eventually enter the negative $D$ region, however typically its value is low. In any case, the tail behavior is not expected to be captured by approximations based on just two cumulants.

%%% %
%%%% TABLE 4
\begin{table*}
\caption{Mean and variance of squared Bures distance between a pure state and a random state: Comparison between RMT and spin-chain results. In each case, along with the dimensions $n,m$, the number of spin sites $L$ and the $S^z$ sector used for obtaining the reduced density matrix from the spin-chain Hamiltonian are also indicated.}
\begin{tabular}{|c|c|c|c|c|c|c|c|c|c|c|c|}
\hline
\multirow{2}{*}{$n$} & \multirow{2}{*}{$m$} & \multirow{2}{*}{$L$} &  \multirow{2}{*}{$S^z$} & \multicolumn{4}{c|}{$\langle D \rangle $} & %
    \multicolumn{4}{c|}{Var$(D)$} \\
\cline{5-12}
& & & & RMT & Spin Chain & Abs. Diff. & Rel. Diff.(\%) & RMT & Spin Chain & Abs. Diff. & Rel. Diff.(\%) \\
\hline
12 & 21 & 10 & 0 & $1.4258 $& $1.4270$ & $1.20\times 10^{-3}$ & $8.42\times 10^{-2}$& $3.62\times 10^{-3}$ & $4.34\times 10^{-3}$ & $7.2\times 10^{-4}$& $19.9$\\
\hline
14 & 18 & 10 & 0 & $1.4689 $ & 1.4704 & $1.50\times 10^{-3}$ & $1.02\times 10^{-1} $ & $3.66 \times 10^{-3}$ & $4.32 \times 10^{-3}$ & $6.6\times 10^{-4}$ & $18.0$   \\
\hline 
15 & 22 & 11 & $\frac{3}{2}$ & $1.4863$ & 1.4732 & $1.31\times 10^{-2}$ & $8.8\times 10^{-1}$ & $2.81 \times 10^{-3}$ & $3.21 \times 10^{-3}$ & $4.0\times 10^{-4}$ & $14.2$ \\
\hline
\end{tabular}
\label{SCps}
\end{table*}

%%% %
%%%% TABLE 5
\begin{table*}
\caption{Mean and variance of squared Bures distance between a maximally mixed state and a random state: Comparison between RMT and spin-chain results.}
\begin{tabular}{|c|c|c|c|c|c|c|c|c|c|c|c|}
\hline
\multirow{2}{*}{$n$} & \multirow{2}{*}{$m$} & \multirow{2}{*}{$L$} &  \multirow{2}{*}{$S^z$} & \multicolumn{4}{c|}{$\langle D \rangle $} & %
    \multicolumn{4}{c|}{Var$(D)$} \\
\cline{5-12}
& & & & RMT & Spin Chain & Abs. Diff. & Rel. Diff.(\%) & RMT & Spin Chain & Abs. Diff. & Rel. Diff.(\%) \\
\hline
12 & 21 & 10 & 0 & $1.545 \times 10^{-1}$& $1.599 \times 10^{-1}$ & $5.4\times 10^{-3}$ & $3.5$& $2.5\times 10^{-4}$ & $2.7\times 10^{-4}$ & $2\times 10^{-5}$& $8.0$\\
\hline
14 & 18 & 10 & 0 & $2.196 \times 10^{-1}$ & $2.262\times 10^{-1} $ & $6.6\times 10^{-3}$ & $3.0$  & $3.2\times 10^{-4}$ & $3.4\times 10^{-4}$ & $2\times 10^{-5}$ & $6.2$\\
\hline 
21 & 22 & 11 & $\frac{1}{2}$ & $2.833 \times 10^{-1}$ & $2.873 \times 10^{-1}$ & $4.0 \times 10^{-3}$ & $1.4$ & $1.95 \times 10^{-4}$ & $2.0\times 10^{-4}$ & $5\times 10^{-6}$ & 2.6 \\
\hline
\end{tabular}
\label{SCmms}
\end{table*}
%%% %
%%%% TABLE 6
\begin{table*}
\caption{Mean and variance of squared Bures distance between two random states: Comparison between RMT and spin-chain results. The two reduced states in the spin-chain case correspond to the bipartitions $(n,m_1)$ and $(n,m_2)$, respectively, as shown in the first three columns and have been obtained from the two $S^z$ sectors indicated in the fifth column. The number of spin sites $L$ is mentioned in the fourth column.}
\begin{tabular}{|c|c|c|c|c|c|c|c|c|c|c|c|c|}
\hline
\multirow{2}{*}{$n$} & \multirow{2}{*}{$m_1$} & \multirow{2}{*}{$m_2$} & \multirow{2}{*}{$L$} &  \multirow{2}{*}{$S^z$} & \multicolumn{4}{c|}{$\langle D \rangle $} & %
    \multicolumn{4}{c|}{Var$(D)$} \\
\cline{6-13}
& & & & & RMT & Spin Chain & Abs. Diff. & Rel. Diff.(\%) & RMT & Spin Chain & Abs. Diff. & Rel. Diff.(\%) \\
\hline
7 & 36 & 30 & 10 & 0,1 & $1.090 \times 10^{-1}$& $1.113 \times 10^{-1}$ & $2.3\times 10^{-3}$ & $2.1$& $4.7\times 10^{-4}$ & $5.0\times 10^{-4}$ & $3\times 10^{-5}$& $6.4$ \\
\hline
14 & 18 & 15 & 10 & 0,1 & $4.276\times 10^{-1}$ & $4.362\times 10^{-1}$ & $8.6\times 10^{-3}$ & $2.0$ & $1.46\times 10^{-3}$ & $1.54\times 10^{-3}$ & $8\times 10^{-5}$ & $5.5$\\
\hline 
11 & 42 & 30 & 11 & $\frac{1}{2},\frac{3}{2}$ & $1.563\times 10^{-1}$ & $1.608\times 10^{-1}$ & $4.5\times 10^{-3}$ & $2.9$  & $3.7\times 10^{-4}$ & $3.9\times 10^{-4}$ & $2\times 10^{-5}$ & $5.4$  \\
\hline
\end{tabular}
\label{SC2r}
\end{table*}
%%% %

%%%%%%
\section{Application in quantum chaotic and complex systems}
\label{sec4}
%%%%%%

In this section, we obtain within the bipartite framework, the statistics of Bures distance using reduced density matrices calculated using the paradigmatic coupled kicked top quantum chaotic system and a correlated spin-chain system in a random magnetic field. These results are compared with the corresponding RMT based analytical results derived in the preceding sections.

%%%%%%
\subsection{Quantum chaotic coupled kicked top system}
%%%%%%

The Hamiltonian for the coupled kicked top system is given by~\cite{MS1999,BL2002}
\begin{equation}
\label{ham}
H=H_1\otimes\mathds{1}_{m}+\mathds{1}_{n}\otimes H_2+H_{12}.
\end{equation}
Here, for the individual tops the Hamiltonian is~\cite{HKS1987,H2010},
\begin{equation}
\label{indham}
H_r=\frac{\pi}{2}J_{y_r}+\frac{\kappa_r}{2j_r}J_{z_r}^{2}\sum_{\nu=-\infty}^{\infty}\delta(t-\nu), r=1,2,
\end{equation}
and the interaction term $H_{12}$ is given by,
\begin{equation}
\label{intr}
H_{12}=\frac{\epsilon}{\sqrt{j_{1}j_{2}}}(J_{z1}\otimes J_{z2})\sum_{\nu=-\infty}^{\infty}\delta(t-\nu).
\end{equation}
The Hamiltonians for individual top $H_1$ and $H_2$ are associated with $n (=2j_1+1)$-dimensional and $m (=2j_2+1)$-dimensional Hilbert spaces $\mathcal{H}^{(n)}$ and $\mathcal{H}^{(m)}$ respectively. The full Hamiltonian for the coupled kicked tops ($H$) correspond to the Hilbert space $\mathcal{H}^{(nm)}=\mathcal{H}^{(n)}\otimes \mathcal{H}^{(m)}$. The angular momentum operators for the $r$th top are $J_{x_r},J_{y_r},J_{z_r}$ and $j_r$ is the quantum number associated with the operator $J_{r}^{2}=J_{x_r}^2+J_{y_r}^2+J_{z_r}^2$. The kick strengths in the individual tops are decided by the stochasticity parameters $\kappa_1$ and $\kappa_2$, respectively. Further, the coupling between the two tops is provided by the parameter $\epsilon$.

The Floquet operator corresponding to the above Hamiltonian is given by~\cite{MS1999,BL2002},
\begin{equation}
\label{floq}
U=U_{12}(U_1\otimes U_2),
\end{equation}
where,
\begin{equation}
U_{r}=\mathrm{exp}\left(-\frac{\imath\kappa_r}{2j_r}J_{z_r}^{2}\right)\mathrm{exp}\left(-\frac{\imath\pi}{2}J_{y_r}\right), ~r=1,2,
\end{equation}
\begin{equation}
U_{12}=\mathrm{exp}\left(-\frac{\imath\epsilon}{\sqrt{j_1j_2}}J_{z_1}\otimes J_{z_2}\right),
\end{equation}
with $\imath=\sqrt{-1}$ being the imaginary-number unit. The above Floquet operator time evolves a state vector from immediately after one kick to immediately after the next. An iterative scheme is implemented to obtain a state $|\psi(\nu)\rangle$ starting from an initial state $|\psi(0)\rangle$ with the aid of $U$ using the relation $|\psi(\nu)\rangle=\mathcal{F}|\psi(\nu-1)\rangle$. The initial state is taken as the product state comprising directed angular momentum states associated with the two tops~\cite{MS1999,BL2002}. Once the iterations are beyond the transient regime, we construct an ensemble of reduced density matrices by considering several $\nu$ values and partial tracing over one of the tops which corresponds to the Hilbert-space dimension $m$, viz., $\rho(\nu)=\mathrm{tr}_{m}(|\psi(\nu)\rangle\langle\psi(\nu)|)$. These reduced density matrices have been shown to be well described by the Hilbert-Schmidt measure for adequately tuned values of the parameters $\kappa_1, \kappa_2$ and $\epsilon$~\cite{KSA2017,K2020,LAK2021,BL2002,FMT2003,BL2004,DK2004,TMD2008,KAT2013,SK2021}.

For our analysis, we simulated 80000 reduced density matrices and obtained mean and variance of their squared Bures distance from a fixed matrix which we choose to be a pure state and then a maximally mixed state.  To simulate the distance between two random density matrices we consider two independent coupled kicked tops (say, $A$ and $B$), so as to realize different $m$ values, viz., $m_1=2j_{2}^{A}+1$ and $m_2=2j_{2}^{B}+1$. Here we use the notations $j_{2}^{A}$ and $j_{2}^{B}$ to represent the $j_2$ values for the pair of coupled kicked tops considered. The common dimension of the two reduced states, $n$, is determined by the common Hilbert-space dimension, $n=2j_{1}^{A}+1=2j_{1}^{B}+1$. The comparison between RMT analytical results and those obtained from coupled-kicked-top simulation for the mean and variance of squared Bures distance are compiled in Tables.~\ref{CKTps}-\ref{CKT2r}. We have shown the results for three different choices of $\kappa_1,\kappa_2$ and $\epsilon$ in each case and find good agreement.

We also illustrate the comparison between the approximated PDF of squared Bures distance and coupled-kicked-tops simulation results in Fig.~\ref{CKTpdf}. A very good agreement can be observed in all the plots.

%%%%FIG5
\begin{figure*}
\advance\leftskip-1cm
\advance\rightskip-3cm
\includegraphics[width=1.0\linewidth]{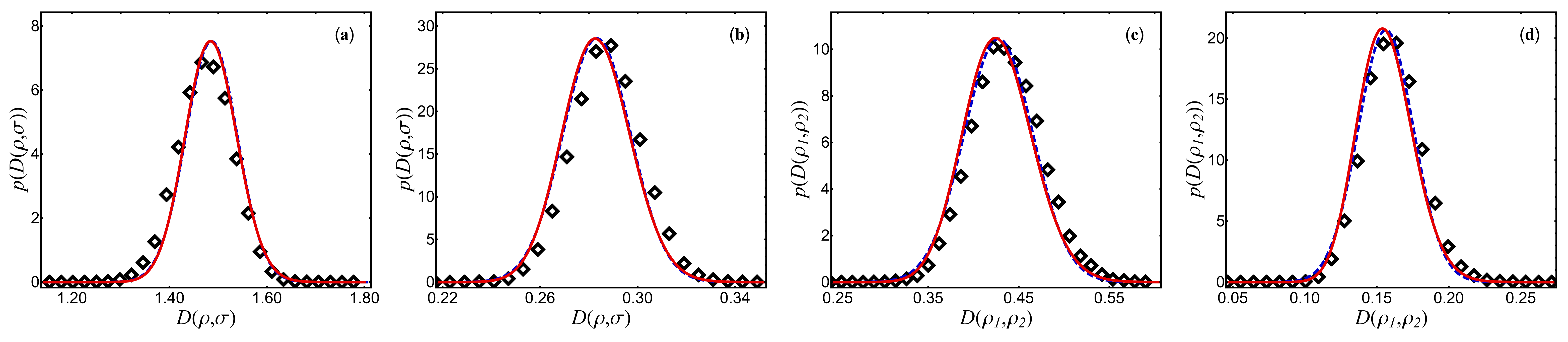}
\centering
\caption{Comparison of approximated PDF and spin-chain simulations between (a) a random density matrix $\rho$ and a pure state $\sigma$ with $(n, m)=(15, 22)$; (b) a random density matrix $\rho$ and a maximally mixed state $\sigma$ with $(n, m)=(21, 22)$; and (c), (d) two random density matrices $\rho_1$ and $\rho_2$ with $(n, m_1, m_2)=(14,18,15)$ and $(11,42,30)$ respectively. The solid lines represent gamma distribution approximation and the dashed lines are using Gaussian distribution approximation. The symbols (diamonds) have been obtained from spin-chain simulations.}
\label{SCpdf}
\end{figure*}
%%%%%

%%%%%%
\subsection{Correlated spin chain in random magnetic field}
%%%%%%

We consider the following one-dimensional spin-1/2 Hamiltonian comprising $L$ sites~\cite{KKG2022},
\begin{equation}
H=\sum_{j=1}^{L}[J\bS_j\cdot \bS_{j+1}+h_j S_j^z+ K \bS_j\cdot(\bS_{j+1}\times \bS_{j+2})],
\end{equation}
where $\bS_j$ represents the spin operator at site $j$, and $S_j^z$ gives the corresponding $z$ component. In the above expression periodic boundary condition is assumed, so that $\bS_{L+k}=\bS_k$.
With the summation considered over the sites, the first term in the above equation corresponds to the isotropic Heisenberg contribution (XXX model), the second term represents the coupling of the spin system to a random magnetic field, and the third term gives the three-site scalar spin-chirality contribution. Correspondingly, $J$ is the nearest-neighbor exchange interaction, $h_j$ gives the strength of site-dependent random (Gaussian, with mean 0 and variance $h^2$) magnetic field along the $z$ direction, and $K$ is the coupling constant for the spin-chirality term. The above Hamiltonian has been used in Ref.~\cite{KKG2022} to study various symmetry crossovers by tuning the strengths of the coupling parameters. It is observed that the total $z$-component of spin, $S^z=\sum_{j=1}^N S_j^z$ commutes with $H$ and therefore when written in the corresponding basis, it block diagonalizes $H$. With all the terms present, both conventional and unconventional time reversal symmetries are broken and then one expects these diagonal blocks, when of considerable sizes, to exhibit spectral statistics matching that of the Gaussian unitary ensemble (GUE) of random matrix theory. We consider these blocks to construct ensembles of random reduced states using the bipartite formalism by considering roughly one-fourth of the middle eigenstates to generate about 50000 squared Bures distance data points. The statistics of Bures distance calculated using these random reduced states are then compared with the RMT analytical results.

For our analyses, we consider $L=10$ and $L=11$ for which the overall Hamiltonian sizes are $2^{10}=1024$ and $2^{11}=2048$, respectively. The parameters are set to $J=1, h=0.35$ and $K=0.7$. In the $L=10$ case, the largest block corresponding to $S^z=0$ sector is of size $252$ and the corresponding spectral fluctuations match well with GUE. We use this with bipartitions $(n,m)=(12,21)$ and $(14,18)$. In the $L=11$ case, the largest block corresponds to $S^z=1/2$ sector and is of size $462$.  In this case, we consider the bipartitions $(n,m)=(11,42), (14,33)$ and $(21,22)$. The next largest block is associated with $S^z=3/2$ sector and is of size $330$. For this, we consider the bipartitions $(11,30)$ and $(15,22)$. The spectral fluctuations for these blocks also show very good agreement with GUE.

In Tables~\ref{SCps}-\ref{SC2r}, we compile the squared Bures distance mean and variance datasets obtained from the spin-chain simulation using the above choices of dimensions, along with the corresponding RMT predictions. Moreover, in Fig.~\ref{SCpdf}, we show the comparison of the corresponding PDFs. We can see from these comparisons that overall the agreement is satisfactory, although not as close as in the case of coupled kicked top system. This may be attributed to the observations that even the midspectrum eigenstates of complex many-body systems do show some deviation from the behavior of random states~\cite{Huang2021,HMK2022,Huang2022}.

%%%%%%%%%
\section{Summary and outlook}
\label{sec5}
%%%%%%%%%
In this work, we have derived exact analytical expressions for the average fidelity and the variance of the squared Bures distance between a pair of density matrices, where one or both are random and taken from the Hilbert-Schmidt distribution. With the help of mean and variance, we also proposed a gamma-distribution-based approximation for the PDF of the square Bures distance. We verified our analytical results using Monte Carlo simulations of relevant random matrix models. We also compared our analytical results with the simulation results obtained using coupled-kicked-tops system and correlated spin-chain in a random magnetic field. We found the agreement to be very good. 

It would be interesting to compare the exact results derived here in other quantum chaotic and complex systems and thereby see the extent of agreement with RMT results. As far as analytical exploration is concerned, one may attempt to derive exact results for other distance measures which are useful in various contexts within quantum information theory, such as trace distance and Hellinger distance. Moreover, concerning the random density matrices, one may examine those from Bures-Hall measure which is also known to be an important class of distribution besides the Hilbert-Schmidt measure~\cite{SK2021,BS2001,SZ2003,FK2016,SK2019,Wei2020a,Wei2020b}.

\appendix

%%%%%%%%%%%%%%%%%%
\section{Proof of Eq.~\eqref{2nd}}
\label{SecApp1}

We derive here Eq.~\eqref{2nd} by evaluating the various averages involved in it. We start with,
\begin{align}
\int_{-\infty}^\infty x R_1(x)dx=\int_{-\infty}^\infty x \frac{1}{\Delta(\{a\})}\sum_{i=1}^N \det[g_{j,k}^{(i)}(x,x)]_{j,k=1}^n.
\end{align}
In the matrix $[g_{j,k}^{(i)}(x,x)]_{j,k=1}^n$ the $x$ variable appears only in one ($i$th) column. We can therefore push the $x$-integral in this column to perform the integral over the determinant. This step is readily justified by expansion of the determinant, application of the integral on individual terms and then recasting of the resulting expansion again as a determinant. Euler's gamma function integral gives the elements of this column as
\begin{align}
\int_{-\infty}^\infty \frac{a_j^m x^{m-i+1} e^{-a_jx} \Theta(x)}{\Gamma(m-i+1)}dx=a_j^{i-2}(m-i+1),
\end{align}
while rest of the $x$-independent columns remain unaltered, and hence we are led to Eq.~\eqref{1st}.

In the same manner, we obtain 
\begin{align}
\int_{-\infty}^\infty x^{1/2} R_1(x)dx= \frac{1}{\Delta(\{a\})}\sum_{i=1}^N \det[\eta_{j,k}^{i,i}]_{j,k=1}^n,
\end{align}
where $\eta_{j,k}^{i,i}$ is as defined in Eq.~\eqref{eta}. Now the first term in Eq.~\eqref{lambdajkhalf} involves the square of the above expression, i.e.,
\begin{align}
\label{1stS}
\nonumber
&\int_{-\infty}^\infty x_1^{1/2} R_1(x_1)dx_1 \int_{-\infty}^\infty x_2^{1/2} R_2(x_2)dx_2\\
&= \frac{1}{\Delta^2(\{a\})}\sum_{i,l=1}^N \det[\eta_{j,k}^{i,i}]_{j,k=1}^n\det[\eta_{j,k}^{l,l}]_{j,k=1}^n,
\end{align}

To obtain the second term in Eq.~\eqref{lambdajkhalf}, we notice that the kernel $S(x_1,x_2)$ can also be written by pulling the factor $x_1^{n-i}\Theta(x_1)$ appearing in the $i$th column matrix elements as
\begin{align}
S(x_1,x_2)= \frac{1}{\Delta(\{a\})}\sum_{i=1}^{n}x_1^{n-i}\Theta(x_1)\det[\hat{g}_{j,k}^{(i)}(x_2)]_{j,k=1}^{n},
\end{align}
where,
\begin{align}
\widehat{g}^{(i)}_{j,k}(x)=\begin{cases}\frac{a_j^m x_2^{m-n}~e^{-a_j x_2}\Theta(x_2)}{\Gamma(m-i+1)} & k=i,\\
 a_{j}^{k-1}, &  k\ne i.
\end{cases}
\end{align}
Therefore, in the expression of $x_1^{1/2} x_2^{1/2} S(x_1,x_2)S(x_2,x_1)$, which involves double sum (over $i$ and $l$), the $x_1$ and $x_2$ dependences factorize. The factor $x_1^{n-i+1/2}\Theta(x_1)$ is now pushed in the $l$th column of the matrix contained in the determinant within $S(x_2,x_1)$. Similarly, the factor $x_2^{n-l+1/2}\Theta(x_2)$ in inserted in the $i$th column of the matrix contained in $S(x_1,x_2)$. The integrals over $x_1$ and $x_2$ can now be performed by focussing on these two columns. This results in the expression,
\begin{align}
\label{2ndS}
\nonumber
&\int_{-\infty}^\infty \int_{-\infty}^\infty x_{1}^{1/2} x_{2}^{1/2}S(x_1,x_2)S(x_2,x_1)dx_1  dx_2\\
&= \frac{1}{\Delta^2(\{a\})}\sum_{i,l=1}^N \det[\eta_{j,k}^{i,l}]_{j,k=1}^n\det[\eta_{j,k}^{l,i}]_{j,k=1}^n.
\end{align}
The mixing of $i$ and $l$ indices should be noted here. Now, we subtract Eq.~\eqref{2ndS} from Eq.~\eqref{1stS}. This leads to the cancellation of $i=l$ terms in the two sums and we are left with
\begin{align}
\nonumber
&\left\langle\sum_{{j,k=1}\atop{(j\ne k)}}^{n}\lambda_{j}^{1/2}\lambda_{k}^{1/2}\right\rangle=\frac{1}{nm\,\Delta^2(\{a\})}~~~~~~~\\
\nonumber
&\times\sum_{{i,l=1}\atop{(i\ne l)}}^{n}\Big(\det[\eta^{i,i}_{j,k}]_{j,k=1}^{n}\det[\eta^{l,l}_{j,k}]_{j,k=1}^{n}\\
&~~~~~~ -\det[\eta^{i,l}_{j,k}]_{j,k=1}^{n} \det[\eta^{l,i}_{j,k}]_{j,k=1}^{n}\Big).
\end{align}
Next, we split the $i\ne l$ sum into $i<l$ and $l>i$, followed by an interchange the indices $i$ and $l$ in the second. Owing to the symmetry of the expressions in the two indices, we eventually obtain Eq.~\eqref{2nd}.

%%%%%%%%%%%%%%%%%%
\section{Proof of Eq.~\eqref{mmixed}}
\label{SecApp2}

The correlation functions $\widehat{R}_1(u)$ and $\widehat{R}_2(u_1,u_2)$ for the eigenvalues $\{u_i\}$ of Wishart-Laguerre ensemble are given by
\begin{align}
\label{R1R2L}
&\widehat{R}_1(u)=\widehat{S}(u,u),\\
&\widehat{R}_2(u_1,u_2)=\widehat{R}_1(u_1)\widehat{R}_1(u_2)-\widehat{S}(u_1,u_2)\widehat{S}(u_2,u_1),
\end{align}
where the kernel $\widehat{S}(u_1,u_2)$ is expressible in terms of associated Laguerre polynomials $L_j^{(b)}(u)$ as,
\begin{align}
\label{S12L}
\nonumber
&\widehat{S}(u_1,u_2)=u_1^{(m-n)/2}e^{-u_1/2}u_2^{(m-n)/2}e^{-u_2/2}\\
&\times\sum_{j=0}^{n-1}\frac{j!}{(m-n+j)!}L_j^{(m-n)}(u_1)L_j^{(m-n)}(u_2) \Theta(u_1)\Theta(u_2).
\end{align}
We now calculate various averages required to derive the expression of mean fidelity when $\sigma$ is a maximally mixed state. The eigenvalues $x_j$ of $\sqrt{\sigma}W\sqrt{\sigma}$ in these case are given by $x_j=u_j/n$. Therefore, the average required in Eq.~\eqref{R1avg}, i.e., $\langle \sum_{j=1}^n \lambda_j\rangle$ can be expressed in terms of integral over $u$ as, $(n^2 m)^{-1}\int_{-\infty}^\infty u\widehat{R}_1(u)du$. Now, this integral can be performed by plugging in the expression of $\widehat{R}_1(u)$ in terms of Laguerre polynomial. However, it is also seen that $\int_{-\infty}^\infty u\widehat{R}_1(u)du=\langle \tr W\rangle_W$ which equals $mn$, following the discussion below Eq.~\eqref{mff1}. Hence, $\langle \sum_{j=1}^n \lambda_j\rangle=(n^2m)^{-1}(mn)=n^{-1}.$

The second average, as in Eq.~\eqref{lambdajkhalf}, can be evaluated using the integrals appearing below,
\begin{align}
\label{secondav}
\nonumber
&\left\langle\sum_{j,k=1\atop (j\ne k)}^n\lambda_j^{1/2}\lambda_k^{1/2}\right\rangle\\
\nonumber
&=\frac{1}{n^2m}\int_{-\infty}^\infty \int_{-\infty}^\infty u_1^{1/2}u_2^{1/2}\widehat{R}_2(u_1,u_2)du_1du_2\\
\nonumber
&=\frac{1}{n^2m}\int_{-\infty}^\infty u_1 ^{1/2} \widehat{R}_1(u_1)du_1\int_{-\infty}^\infty u_2^{1/2} \widehat{R}_1(u_2) du_2\\
&-\frac{1}{n^2m}\int_{-\infty}^\infty \int_{-\infty}^\infty u_1^{1/2}u_2^{1/2}\widehat{S}(u_1,u_2)\widehat{S}(u_2,u_1)du_1du_2.
\end{align}
We first examine the integral $\int_{-\infty}^\infty u ^{1/2} \widehat{R}_1(u)du$. This can be evaluated by plugging in the expression of $\widehat{R}_1(u)$ in terms of associated Laguerre polynomials using the following identity due to Schr\"odinger~\cite{Schro1926}:
\begin{align}
\label{Schro}
\nonumber
&\int_0^\infty z^q e^{-z} L_j^{(a)}(z)L_k^{(b)}(z)dz\\
&=(-1)^{j+k}\sum_{r=0}^{\min(j,k)}\binom{q-a}{j-r}\binom{q-b}{k-r}\frac{\Gamma(q+r+1)}{r!};~~r>-1.
\end{align}
We obtain,
\begin{align}
\nonumber
&\int_{-\infty}^\infty u ^{1/2} \widehat{R}_1(u)du\\
&=\sum_{j=0}^{n-1}\frac{j!}{(m-n+j)!}\sum_{r=0}^j \binom{\tfrac{1}{2}}{j-r}^2 \frac{\Gamma(m-n+r+\tfrac{3}{2})}{r!}.
\end{align}
Next, using the summation formula obtained using Mathematica~\cite{WolframMtmk},
\begin{align}
\label{Mtmk}
\nonumber
\sum_{r=0}^j \binom{\tfrac{1}{2}}{j-r}\binom{\tfrac{1}{2}}{k-r}\frac{\Gamma(m-n+r+\tfrac{3}{2})}{r!}=\binom{\tfrac{1}{2}}{j} \binom{\tfrac{1}{2}}{k}\\
\times \Gamma(m-n+\tfrac{3}{2})\,_3F_2(-j,-k,m-n+\tfrac{3}{2}; \tfrac{3}{2}-j,\tfrac{3}{2}-k;1),
\end{align}
we arrive at
\begin{align}
\label{sumform1}
\nonumber
&\int_{-\infty}^\infty u^{1/2} \widehat{R}_1(u)du\\
\nonumber
&=\sum_{j=0}^{n-1}\frac{j!\,  \Gamma(m-n+\tfrac{3}{2})}{(m-n+j)!}\binom{\tfrac{1}{2}}{j}^2\\
&\times\,_3F_2(-j,-j,m-n+\tfrac{3}{2}; \tfrac{3}{2}-j,\tfrac{3}{2}-j;1).
\end{align}
Interestingly, in Ref.~\cite{MS2011}, another expression for the above average has been obtained,
\begin{align}
\label{sumform2}
\int_{-\infty}^\infty u^{1/2} \widehat{R}_1(u)du=\sum_{j=0}^{n-1}\binom{\tfrac{1}{2}}{j}\binom{\tfrac{1}{2}}{j+1}\frac{(m)_{1/2-j} }{(n+1)_{-j-1}}.
\end{align}
It may be noted that the summands in Eqs.~\eqref{sumform1} and~\eqref{sumform2} are not equal.
Now, using~\eqref{sumform1}, the first term in Eq.~\eqref{secondav} can be written as,
\begin{align}
\label{firstsum}
\nonumber
&\int_{-\infty}^\infty u_1^{1/2} \widehat{R}_1(u_1)du_1\int_{-\infty}^\infty u_2^{1/2} \widehat{R}_1(u_2)du_2\\
\nonumber
&=\sum_{j=0}^{n-1}\sum_{k=0}^{n-1}\frac{j!k!\,  \Gamma(m-n+\tfrac{3}{2})^2}{(m-n+j)!(m-n+k)!}\binom{\tfrac{1}{2}}{j}^2\binom{\tfrac{1}{2}}{k}^2\\
\nonumber
&\times\,_3F_2(-j,-j,m-n+\tfrac{3}{2}; \tfrac{3}{2}-j,\tfrac{3}{2}-j;1)\\
&\times\,_3F_2(-k,-k,m-n+\tfrac{3}{2}; \tfrac{3}{2}-k,\tfrac{3}{2}-k;1).
\end{align}

Next, we evaluate the second term in Eq.~\eqref{secondav}. Using again the kernel expression~\eqref{S12L} and the integration formula~\eqref{Schro}, we obtain
\begin{align}
\nonumber
&\int_{-\infty}^\infty \int_{-\infty}^\infty u_1^{1/2}u_2^{1/2}\widehat{S}(u_1,u_2)\widehat{S}(u_2,u_1)du_1du_2\\
\nonumber
&=\sum_{j=0}^{n-1}\sum_{k=0}^{n-1} \frac{j!k!}{(m-n+j)!(m-n+k)!}\\
\nonumber
&\times\sum_{i=0}^{\min(j,k)}\binom{\tfrac{1}{2}}{j-i}\binom{\tfrac{1}{2}}{k-i}\frac{\Gamma(m-n+\tfrac{3}{2}+i)}{i!}\\
&\times\sum_{l=0}^{\min(j,k)}\binom{\tfrac{1}{2}}{j-l}\binom{\tfrac{1}{2}}{k-l}\frac{\Gamma(m-n+\tfrac{3}{2}+l)}{l!}.
\end{align}
Now, noticing that the summand in the above expression is symmetric in $j$ and $k$, we split the sum into $j=k$ and $j<k$ terms and then use the result in Eq.~\eqref{Mtmk}. This gives,
\begin{align}
\label{secondsum}
\nonumber
&\int_{-\infty}^\infty \int_{-\infty}^\infty u_1^{1/2}u_2^{1/2}\widehat{S}(u_1,u_2)\widehat{S}(u_2,u_1)du_1du_2\\
\nonumber
&=\sum_{j=0}^{n-1} \left(\frac{j!}{(m-n+j)!}\right)^2\binom{\tfrac{1}{2}}{j}^4\Gamma^2(m-n+\tfrac{3}{2})\\
\nonumber
&\times[\,_3F_2(-j,-j,m-n+\tfrac{3}{2}; \tfrac{3}{2}-j,\tfrac{3}{2}-j;1)]^2\\
\nonumber
&+ 2\sum_{1\le j<k\le n} \frac{j!k!~\Gamma^2(m-n+\tfrac{3}{2})}{(m-n+j)!(m-n+k)!}\binom{\tfrac{1}{2}}{j}^2\binom{\tfrac{1}{2}}{k}^2\\
&\times [\,_3F_2(-j,-k,m-n+\tfrac{3}{2}; \tfrac{3}{2}-j,\tfrac{3}{2}-k;1)]^2.
\end{align}
In the final step, we split Eq.~\eqref{firstsum} into $j=k$ and $j<k$ terms, and subtract~\eqref{secondsum} from it. The cancellation of $j=k$ term takes place and finally leads us to Eq.~\eqref{mmixed}.

%%%%%%%%%%%%%%%%%%
\section{Proof of Eq.~\eqref{mf2}}
\label{SecApp3}

The following integral identity will be useful in our calculation below~\cite{PBM1990},
\begin{align}
\label{intiden}
\nonumber
&\int_{-\infty}^\infty y^{k+1/2} G^{2,1}_{1,3} \left(
\begin{matrix}
-j\\
v_{1},v_{2};0
\end{matrix} \bigg| y\right)\,\Theta(y)\,dy\\
&=\frac{\Gamma(k+v_1+\tfrac{3}{2})\Gamma(k+v_2+\tfrac{3}{2})\Gamma(j-k-\tfrac{1}{2})}{\Gamma(-k-\tfrac{1}{2})}.
\end{align}
We focus on the first term of Eq.~\eqref{mff2} and evaluate $\int_{-\infty}^\infty y_1^{1/2} \tR_1(y_1)dy_1$. Use of Eqs.~\eqref{Rt1y} and~\eqref{MGint}, along with the integral in Eq.~\eqref{intiden} above, gives
 \begin{align}
 \label{secav}
 \nonumber
& \int_{-\infty}^\infty y_1^{1/2} \tR_1(y_1)dy_1\\
\nonumber
&=\sum_{j=0}^{n-1}\sum_{k=0}^j \frac{(-1)^k}{(j-k)!}\frac{\Gamma(k+v_1+\tfrac{3}{2})\Gamma(k+v_2+\tfrac{3}{2})\Gamma(j-k-\tfrac{1}{2})}{k!(k+v_1)!\,(k+v_2)!\,\Gamma(-k-\tfrac{1}{2})}\\
&=\sum_{k=0}^{n-1}\sum_{j=0}^{n-k-1}\frac{(-1)^k}{j!}\frac{\Gamma(k+v_1+\tfrac{3}{2})\Gamma(k+v_2+\tfrac{3}{2})\Gamma(j-\tfrac{1}{2})}{k!(k+v_1)!\,(k+v_2)!\,\Gamma(-k-\tfrac{1}{2})}.
 \end{align}
In arriving at the second step above, we have used the result
\begin{equation}
\sum_{j=0}^{n-1}\sum_{k=0}^j b_{k,j}=\sum_{k=0}^{n-1}\sum_{j=k}^{n-1} b_{k,j}=\sum_{k=0}^{n-1}\sum_{j=0}^{n-k-1} b_{k,j+k},
\label{reorder}
\end{equation}
which is obtained by considering change in the order of summation, followed by a shift in the inner summation index.
The sum over $j$ can be performed in the second expression in the Eq.~\eqref{secav} using the result 
\begin{align}
\label{sumform}
\sum_{j=0}^{c}\frac{\Gamma(j+b)}{j!}=\frac{\Gamma(b+c+1)}{b\,\Gamma(c+1)}.
\end{align}
We further use Euler's reflection formula to write $\Gamma(-k-1/2)=(-1)^{k-1}\pi/\Gamma(k+3/2)$. With these in Eq.~\eqref{secav}, we obtain
 \begin{align}
 \nonumber
& \int_{-\infty}^\infty y_1^{1/2} \tR_1(y_1)dy_1\\
\nonumber
&=\frac{2}{\pi}\sum_{k=0}^{n-1}\frac{\Gamma(k+3/2)\Gamma(k+v_1+\tfrac{3}{2})\Gamma(k+v_2+\tfrac{3}{2})\Gamma(n-k-\tfrac{1}{2})}{k!(k+v_1)!\,(k+v_2)!\,\Gamma(n-k)}\\
&=\frac{2}{\pi}\sum_{k=1}^{n} (k)_{1/2}(k+v_1)_{1/2}(k+v_2)_{1/2}(n-k+1)_{-1/2}.
 \end{align}
As a consequence, 
 \begin{align}
 \label{expr1}
 \nonumber
& \int_{-\infty}^\infty y_1^{1/2} \tR_1(y_1)dy_1\int_{-\infty}^\infty y_2^{1/2} \tR_2(y_2)dy_2\\
\nonumber
&=\frac{4}{\pi^2}\sum_{j,k=1}^{n} (j)_{1/2}(j+v_1)_{1/2}(j+v_2)_{1/2}(n-j+1)_{-1/2}\\
&\times (k)_{1/2}(k+v_1)_{1/2}(k+v_2)_{1/2}(n-k+1)_{-1/2}.
 \end{align}

Now, we evaluate the second term in Eq.~\eqref{mff2}, using Eqs.~\eqref{Sy1y2} and~\eqref{intiden}, We obtain
\begin{align}
\nonumber
&\int_{-\infty}^\infty \int_{-\infty}^\infty y_1^{1/2}y_2^{1/2} \tS(y_1,y_2)\tS(y_2,y_1)dy_1 dy_2\\
\nonumber
&=\sum_{i=0}^{n-1}\sum_{j=0}^i\sum_{l=0}^{n-1}\sum_{k=0}^l \frac{(-1)^j}{j!(j+v_1)!(j+v_2)!(i-j)!} \\
\nonumber
&\times \frac{(-1)^k}{k!(k+v_1)!(k+v_2)!(l-k)!}\\
\nonumber
&\times \frac{\Gamma(j+v_1+3/2)\Gamma(j+v_2+3/2)\Gamma(l-j-1/2)}{\Gamma(-j-1/2)}\\
&\times \frac{\Gamma(k+v_1+3/2)\Gamma(k+v_2+3/2)\Gamma(i-k-1/2)}{\Gamma(-k-1/2)}.
\end{align}
In the next step, we use the identity in Eq.~\eqref{reorder} to interchange the orders of the summation involving $(i,j)$ pair and over $(l,k)$ pair. Afterwards, we perform the summations over $i$ and $l$ using the result in Eq.~\eqref{sumform}.
This leaves us with,
\begin{align}
\nonumber
&\int_{-\infty}^\infty \int_{-\infty}^\infty y_1^{1/2}y_2^{1/2} \tS(y_1,y_2)\tS(y_2,y_1)dy_1 dy_2\\
\nonumber
&=\sum_{j=0}^{n-1}\sum_{k=0}^{n-1} \frac{(-1)^j(-1)^k}{j!(j+v_1)!(j+v_2)k!(k+v_1)!(k+v_2)!} \\
\nonumber
&\times \frac{\Gamma(j+v_1+\tfrac{3}{2})\Gamma(j+v_2+\tfrac{3}{2})\Gamma(k+v_1+\tfrac{3}{2})\Gamma(k+v_2+\tfrac{3}{2})}{\Gamma(-j-1/2)\Gamma(-k-1/2)}\\
&\times \frac{\Gamma(n-j-\tfrac{1}{2})\Gamma(n-k-\tfrac{1}{2})}{(j-k-\tfrac{1}{2})(k-j-\tfrac{1}{2})\Gamma(n-j)\Gamma(n-k)}.
\end{align}
We now use Euler's reflection formula for $\Gamma(-j-1/2)$ and $\Gamma(-k-1/2)$, shift the summation indices to run from $1$ to $n$, and express ratio of Gamma functions in terms of Pochhammer symbols. This gives,
\begin{align}
\label{expr2}
\nonumber
&\int_{-\infty}^\infty \int_{-\infty}^\infty y_1^{1/2}y_2^{1/2} \tS(y_1,y_2)\tS(y_2,y_1)dy_1 dy_2\\
\nonumber
&=\frac{1}{\pi^2}\sum_{j=1}^{n}\sum_{k=1}^{n} \frac{1}{1/4-(j-k)^2} \\
\nonumber
&\times (j)_{1/2}(j+v_1)_{1/2}(j+v_2)_{1/2}(n-j+1)_{-1/2}\\
&\times (k)_{1/2}(k+v_1)_{1/2}(k+v_2)_{1/2}(n-k+1)_{-1/2}.
\end{align}

Now, subtracting~\eqref{expr2} from~\eqref{expr1}, we obtain
 \begin{align}
 \nonumber
& \int_{-\infty}^\infty y_1^{1/2} \tR_1(y_1)dy_1\int_{-\infty}^\infty y_2^{1/2} \tR_2(y_2)dy_2\\
&-\int_{-\infty}^\infty \int_{-\infty}^\infty y_1^{1/2}y_2^{1/2} \tS(y_1,y_2)\tS(y_2,y_1)dy_1 dy_2\\
\nonumber
&=\frac{4}{\pi^2}\sum_{j=1}^{n}\sum_{k=1}^{n} \frac{(j-k)^2}{(j-k)^2-1/4} \\
\nonumber
&\times (j)_{1/2}(j+v_1)_{1/2}(j+v_2)_{1/2}(n-j+1)_{-1/2}\\
&\times (k)_{1/2}(k+v_1)_{1/2}(k+v_2)_{1/2}(n-k+1)_{-1/2}.
 \end{align}
Equation~\eqref{mf2} then follows by noticing that the summand in the above expression is symmetric in $j$ and $k$, and that $j=k$ term vanishes.

%\section*{References}


\begin{thebibliography}{199}




% 
\bibitem{NC2000} M. A. Nielsen and I. L. Chuang, {\it Quantum Computation and Quantum Information} (Cambridge University Press, Cambridge, 2000).
%
\bibitem{E2009} E. Desurvire, Classical and Quantum Information Theory: An Introduction for the Telecom Scientist (Cambridge University Press, Cambridge, 2009).
% 
\bibitem{BZ2017} I. Bengtsson and K. \.Zyczkowski, {\it Geometry of Quantum States: an Introduction to Quantum Entanglement}, 2nd ed.(Cambridge University Press, Cambridge, 2017)
%
\bibitem{W2017} M. M. Wilde, {\it Quantum Information Theory}, 2nd ed. (Cambridge University Press, Cambridge, UK, 2017).
%
\bibitem{STM2013} T. Sugiyama, P. S. Turner, and M. Murao, Precision-guaranteed quantum tomography, Phys. Rev. Lett. {\bf 111}, 160406 (2013).
%
\bibitem{KKF2020} E. O. Kiktenko, D. N. Kublikova, and A. K. Fedorov, Estimating the precision for quantum process tomography, Opt. Eng. {\bf 59}, 061614 (2020).
%
\bibitem{ZE2011} H. Zhu and B.-G. Englert, Quantum state tomography with fully symmetric measurements and product measurements, Phys. Rev. A {\bf 84}, 022327 (2011).
%
\bibitem{TBCL2019} V. Tr\'{a}vn\'{i}\v{c}ek, K. Bartkiewicz, A. \v{C}ernoch, and K. Lemr, Experimental measurement of the Hilbert-Schmidt distance between two-qubit states as a means for reducing the complexity of machine learning, Phys. Rev. Lett. {\bf 123}, 260501 (2019).
%
\bibitem{Gao2016} Y. Guo, Non-commutativity measure of quantum discord, Sci. Rep. {\bf 6}, 25241 (2016).
%
\bibitem{VPRK1997} V. Vedral, M. B. Plenio, M. A. Rippin, and P. L. Knight, Quantifying entanglement, Phys.Rev. Lett. {\bf 78}, 2275 (1997).
%
\bibitem{ACSZC2019} A. Arrasmith, L. Cincio, A. T. Sornborger, W. H. Zurek, and P. J. Coles, Variational consistent histories as a hybrid algorithm for quantum foundations, Nat. Commun. {\bf 10}, 3438 (2019).
%
\bibitem{LZ2004} S. Luo and Q. Zhang, Informational distance on quantum-state space, Phys. Rev. A {\bf 69}, 032106 (2004).
%
\bibitem{SO2013} D. Spehner and M. Orszag 2013, Geometric quantum discord with Bures distance, New J. Phys. {\bf 15}, 103001 (2013).
%
\bibitem{WPSW2020} M. Wie\'{s}niak, P. Pandya, O. Sakarya, and B. Woloncewicz, Distance between bound entangled states from unextendible product bases and separable states, Quantum Rep. {\bf 2}, 49 (2020).
%
\bibitem{PSW2020} P. Pandya, O. Sakarya, and M. Wie\'{s}niak, Hilbert-Schmidt distance and entanglement witnessing, Phys. Rev. A {\bf 102}, 012409 (2020).
%
\bibitem{LTOCC2019} R. LaRose, A. Tikku, \'{E}. O'Neel-Judy, L. Cincio, and P. J. Coles, Variational quantum state diagonalization, npj Quantum Inf. {\bf 5}, 57 (2019).
% 
\bibitem{KLPCSC2019} S. Khatri, R. LaRose, A. Poremba, L. Cincio, A. T. Sornborger, and P. J. Coles, Quantum-assisted quantum compiling, Quantum {\bf 3}, 140 (2019).
% 
\bibitem{CPCC2019} M. Cerezo, A. Poremba, L. Cincio, and P. J. Coles, Variational quantum fidelity estimation, Quantum {\bf 4}, 248 (2020).
%
\bibitem{RSI2016} W. Roga, D. Spehner, and F. Illuminati, Geometric measures of quantum correlations: characterization, quantification, and comparison by distances and operations, J. Phys. A: Math. Theor. {\bf 49}, 235301 (2016).
%
\bibitem{MMPZ2008} D. Markham, J. Adam Miszczak, Z. Pucha\l{}a, and Karol \.Zyczkowski, Quantum state discrimination: A geometric approach, Phys. Rev. A {\bf 77}, 042111 (2008).
%
\bibitem{PPZ2016} Z. Pucha\l{}a, \L{}. Pawela, and K. \.Zyczkowski, Distinguishability of generic quantum states, Phys. Rev. A {\bf 93}, 062112 (2016).
%
\bibitem{B1969} D. ~Bures, An extension of Kakutani's theorem on infinite product measures to the tensor product of semifinite w*-algebras, Trans. Am. Math. Soc. {\bf 135}, 199 (1969).
%
\bibitem{U1976} A. Uhlmann, The transition probability in the state space of a $\ast$-algebra, Rep. Math. Phys. {\bf 9}, 273 (1976).
%
\bibitem{U1986} A. Uhlmann, Parallel transport and ``quantum holonomy" along density operators, Rep. Math. Phys. {\bf 24}, 229-240 (1986).
%
\bibitem{J1994} R. Jozsa, Fidelity for mixed quantum states, J. Mod. Opt. {\bf 41} 2315 (1994).
%
\bibitem{W1990} W. K. Wootters, Random quantum states, Found. Phys. {\bf 20}, 1365 (1990).
%
\bibitem{H1998} M. J. W. Hall, Random quantum correlations and density operator distributions, Phys. Lett. A. {\bf 242}, 123 (1998).
%
\bibitem{ZS2001} K. \.Zyczkowski and H.-J. Sommers, Induced measures in the space of mixed quantum states, J. Phys. A: Math. Gen. {\bf 34}, 7111 (2001).
%
\bibitem{SZ2004} H.-J. Sommers and K. \.Zyczkowski, Statistical properties of random density matrices, J. Phys. A: Math. Gen. {\bf 37}, 8457 (2004).
% 
\bibitem{ZPNC2011} K. \.Zyczkowski, K. A. Penson, I. Nechita, and B. Collins, Generating random density matrices, J. Math. Phys. {\bf 52}, 062201 (2011).
%
\bibitem{CN2016} B. Collins and I. Nechita, Random matrix techniques in quantum information theory, J. Math. Phys. {\bf 57}, 015215 (2016).
%
\bibitem{HKS1987} F. Haake, M. Ku\'{s} and R. Scharf, Classical and quantum chaos for a kicked top, Z. Phys. B: Condens. Matter {\bf 65}, 381 (1987).
% 
\bibitem{H2010} F. Haake, {\it Quantum Signatures of Chaos} (Springer, New York, 2010).
%
\bibitem{Lubkin1978} E. Lubkin, Entropy of an $n$-system from its correlation with a $k$-reservoir, J. Math. Phys. {\bf 19}, 1028 (1978).
% 
\bibitem{LP1988} S. Lloyd and H. Pagels, Complexity as thermodynamic depth, Ann. Phys., NY {\bf 188}, 186 (1988).
%
\bibitem{Page1993} D. N. Page, Average entropy of a subsystem, Phys. Rev. Lett. {\bf 71}, 1291 (1993).
%
\bibitem{G2007} O. Giraud, Purity distribution for bipartite random pure states, J. Phys. A: Math. Theor. {\bf 40}, F1053 (2007).
%
\bibitem{MBL2008} S. N. Majumdar, O. Bohigas, and A. Lakshminarayan, Exact minimum eigenvalue distribution of an entangled random pure state, J Stat. Phys. {\bf 131}, 33 (2008).
%
\bibitem{NMV2011} C. Nadal, S. N. Majumdar, and M. Vergassola, Statistical distribution of quantum entanglement for a random bipartite state, J. Stat. Phys. {\bf 142}, 403 (2011).
%
\bibitem{KP2011} S. Kumar and A. Pandey, Entanglement in random pure states: spectral density and average von Neumann entropy, J. Phys. A: Math. Theor. {\bf 44}, 445301 (2011). 
%
\bibitem{VPO2016} P. Vivo, M. P. Pato, and G. Oshanin, Random pure states: Quantifying bipartite entanglement beyond the linear statistics, Phys. Rev. E {\bf 93}, 052106 (2016).
%
\bibitem{KSA2017} S. Kumar, B. Sambasivam, and S. Anand, Smallest eigenvalue density for regular or fixed-trace complex Wishart-Laguerre ensemble and entanglement in coupled kicked tops, J. Phys. A: Math. Theor. {\bf 50}, 345201 (2017).
%
\bibitem{K2019} S. Kumar, Recursion for the smallest eigenvalue density of $\beta$-Wishart-Laguerre ensemble, J. Stat. Phys. {\bf 175}, 126 (2019).
%
\bibitem{FK2019} P. J. Forrester and S. Kumar, Recursion scheme for the largest $\beta$ -Wishart-Laguerre eigenvalue and Landauer conductance in quantum transport, J. Phys. A: Math. Theor. {\bf 52}, 42LT02 (2019).
%
\bibitem{W2020} L. Wei, Skewness of von Neumann entanglement entropy, J. Phys. A: Math. Theor. {\bf 53}, 075302 (2020).
%
\bibitem{B1996} S. L. Braunstein, Geometry of quantum inference, Phys. Lett. A {\bf 219}, 169 (1996).
%
\bibitem{ZS2005} K. \.Zyczkowski and H.-J. Sommers, Average fidelity between random quantum states, Phys. Rev. A {\bf 71}, 032313 (2005).
%
\bibitem{HLW2006} P. Hayden, D. Leung, and A. Winter, Aspects of generic entanglement, Commun. Math. Phys. {\bf 265}, 95 (2006).
% 
\bibitem{M2007} A. Montanaro, On the distinguishability of random quantum states, Commun. Math. Phys. {\bf 273}, 619 (2007).
%
\bibitem{BSZW2016}  K. Bu, U. Singh, L. Zhang, J. Wu, Average distance of random pure states from maximally entangled and coherent states, arXiv:1603.06715.
%
\bibitem{MZB2017} J. Mej\'ia, C. Zapata, and A. Botero, The difference between two random mixed quantum states: exact and asymptotic spectral analysis, J. Phys. A: Math. Theor. {\bf 50}, 025301 (2017).
%
\bibitem{KC2020} S. Kumar and S. Sai Charan, Spectral statistics for the difference of two Wishart matrices, J. Phys. A: Math. Theor. {\bf 53}, 505202 (2020).
%
\bibitem{K2020} S. Kumar, Wishart and random density matrices: Analytical results for the mean-square Hilbert-Schmidt distance, Phys. Rev. A {\bf 102}, 012405 (2020).
%
\bibitem{LAK2021} A. Laha,  A. Aggarwal and S. Kumar,  Random density matrices: Analytical results for mean root fidelity and mean-square Bures distance,  Phys. Rev. A,  {\bf 104},  022438,  (2021).
%
\bibitem{G1963} N. R. Goodman, Statistical analysis based on a certain multivariate complex gaussian distribution (An introduction), Ann. Math. Statist. {\bf 34}, 152 (1963).
% 
\bibitem{ATLV2004} G. Alfano, A. M.~Tulino, A. Lozano and S. Verdu, Capacity of MIMO channels with one-sided correlation,  \emph{Eighth IEEE International Symposium on Spread Spectrum Techniques and Applications - Programme and Book of Abstracts (IEEE Cat. No.04TH8738)}, 2004, pp. 515-519.
%
\bibitem{SMM2006} S. Simon, A. Moustakas, and L. Marinelli, Capacity and character expansions: Moment-generating function and other exact results for MIMO correlated channels, IEEE Trans. Inf. Theory {\bf 52}, 5336 (2006).
%
\bibitem{RKG2010} C. Recher, M. Kieburg, and T. Guhr, Eigenvalue densities of real and complex Wishart correlation matrices, Phys. Rev. Lett. {\bf 105}, 244101 (2010).
%
\bibitem{AIK2013} G. Akemann, J. R. Ipsen, and M. Kieburg, Products of rectangular random matrices: Singular values and progressive scattering, Phys. Rev. E {\bf 88}, 052118 (2013).
%
\bibitem{MS1999} P. A. Miller and S. Sarkar, Signatures of chaos in the entanglement of two coupled quantum kicked tops, Phys. Rev. E {\bf 60}, 1542 (1999).
% 
\bibitem{BL2002} J. N. Bandyopadhyay and A. Lakshminarayan, Testing statistical bounds on entanglement using quantum chaos, Phys. Rev. Lett. {\bf 89}, 060402 (2002).
% 
\bibitem{FMT2003} H. Fujisaki, T. Miyadera and A. Tanaka, Dynamical aspects of quantum entanglement for weakly coupled kicked tops, Phys. Rev. E {\bf 67}, 066201 (2003).
% 
\bibitem{BL2004} J. N. Bandyopadhyay and A. Lakshminarayan, Entanglement production in coupled chaotic systems: Case of the kicked tops, Phys. Rev. E {\bf 69}, 016201 (2004).
% 
\bibitem{DK2004} R. Demkowicz-Dobrza\'nski and M. Ku\'s, Global entangling properties of the coupled kicked tops, Phys. Rev. E {\bf 70}, 066216 (2004).
% 
\bibitem{TMD2008} C. M. Trail, V. Madhok and I. H. Deutsch, Entanglement and the generation of random states in the quantum chaotic dynamics of kicked coupled tops, Phys. Rev. E {\bf 78}, 046211 (2008).
% 
\bibitem{KAT2013} H. Kubotani, S. Adachi and M. Toda, Measuring dynamical randomness of quantum chaos by statistics of Schmidt eigenvalues, Phys. Rev. E {\bf 87}, 062921 (2013).
%
\bibitem{SK2021} A. Sarkar and S. Kumar, Generation of Bures-Hall mixed states using coupled kicked tops, Phys. Rev. A {\bf 103}, 032423 (2021).
%
\bibitem{KKG2022} D. Kundu, S. Kumar, and S. Sen Gupta, Spectral crossovers and universality in quantum spin chains coupled to random fields, Phys. Rev. B {\bf 105}, 014205 (2022).
%
\bibitem{Huang2021} Y. Huang, Universal entanglement of mid-spectrum eigenstates of chaotic local Hamiltonians, Nucl. Phys. B {\bf 966}, 115373 (2021).
%
\bibitem{HMK2022} M. Haque, P. A. McClarty, and I. M. Khaymovich. Entanglement of midspectrum eigenstates of chaotic many-body systems: Reasons for deviation from random ensembles. Phys. Rev. E, {\bf 105} 014109 (2022).
%
\bibitem{Huang2022} Y. Huang, Deviation from maximal entanglement for mid-spectrum eigenstates of local Hamiltonians, arXiv:2202.01173 (2022).
%
\bibitem{BS2001} M. S. Byrd and P. B. Slater, Bures measures over the spaces of two- and three-dimensional density matrices, Phys. Lett. A {\bf 283}, 152 (2001).
%
\bibitem{SZ2003} H.-J. Sommers and K. \.Zyczkowski, Bures volume of the set of mixed quantum states, J. Phys. A: Math. Gen. {\bf 36}, 10083 (2003).
%
\bibitem{FK2016} P. J. Forrester and M. Kieburg, Relating the Bures measure to the Cauchy two-matrix model, Commun. Math. Phys. {\bf 342}, 151 (2016)
%
\bibitem{SK2019} A. Sarkar and S. Kumar, Bures-Hall ensemble: spectral densities and average entropies, J. Phys. A: Math. Theor. {\bf 52}, 295203 (2019).
%
\bibitem{Wei2020a} L. Wei, Proof of Sarkar-Kumar's conjectures on average entanglement entropies over the Bures-Hall ensemble, J. Phys. A: Math. Theor. {\bf 53}, 235203 (2020).
%
\bibitem{Wei2020b} L. Wei, Exact variance of von Neumann entanglement entropy over the Bures-Hall measure, Phys. Rev. E {\bf 102}, 062128 (2020).
%
\bibitem{Schro1926} E. Schr\"odinger, Quantisierung als Eigenwertproblem, Ann. Phys. (Leipzig) {\bf 385}, 437 (1926).
%
\bibitem{MS2011} F. Mezzadri and N. J. Simm, Moments of the transmission eigenvalues, proper delay times, and random matrix theory. I
, J. Math. Phys. 52, 103511 (2011).
%
\bibitem{WolframMtmk} Wolfram Research, Inc., Mathematica, Version 13.1, Champaign, IL (2022).
%
\bibitem{PBM1990} A. P. Prudnikov, Y. A. Brychkov, and O. I. Marichev, {\it Integrals and Series: More Special Functions}, Vol. 3 (Gordon and Breach, London, 1990).



\end{thebibliography}
\end{document}